\documentclass[journal]{IEEEtran}

\ifCLASSINFOpdf
 
\else
 
\fi

\usepackage{cite}
\usepackage{framed}
\usepackage{threeparttable}
\usepackage{tabularx}
\usepackage{diagbox}
\usepackage{fancyhdr}
\usepackage[normalem]{ulem}
\usepackage{hyperref}
\usepackage{amsmath}
\usepackage{amsfonts}
\usepackage{booktabs} % For formal tables
\usepackage{upgreek}
\usepackage{graphicx}
\usepackage{verbatim}
\usepackage{multirow}
\usepackage{array}
\usepackage{subfigure}
\usepackage{float}
\usepackage{CJK}
\usepackage{setspace}
\usepackage[font=footnotesize]{caption}
\usepackage{indentfirst}
\usepackage{dblfloatfix} 
\usepackage{multirow}
\usepackage{algorithmicx,algorithm}
\usepackage{xcolor}
\usepackage{tikz}
\usepackage{color}
\graphicspath{{./Fig/}}

\usepackage[noend]{algpseudocode}

\usepackage[subtle]{savetrees}
\newcommand*\circlednew[2]{\tikz[baseline=(char.base)]{
            \node[shape=circle,fill=black,inner sep=1pt] (char) {\textcolor{#1}{{\footnotesize #2}}};}}

  \DeclareSymbolFont{numbers}{T1}{ptm}{m}{n}
  \SetSymbolFont{numbers}{bold}{T1}{ptm}{bx}{n}
  \DeclareMathSymbol{0}\mathalpha{numbers}{"30}
  \DeclareMathSymbol{1}\mathalpha{numbers}{"31}
  \DeclareMathSymbol{2}\mathalpha{numbers}{"32}
  \DeclareMathSymbol{3}\mathalpha{numbers}{"33}
  \DeclareMathSymbol{4}\mathalpha{numbers}{"34}
  \DeclareMathSymbol{5}\mathalpha{numbers}{"35}
  \DeclareMathSymbol{6}\mathalpha{numbers}{"36}
  \DeclareMathSymbol{7}\mathalpha{numbers}{"37}
  \DeclareMathSymbol{8}\mathalpha{numbers}{"38}
  \DeclareMathSymbol{9}\mathalpha{numbers}{"39}

\begin{document}

% \title{\Huge{A Comprehensive Design Automation Framework for Analog Circuit Parameter
% Optimization}}

% \title{\Huge{AutoSizing:  Domain Knowledge-Infused Deep Learning for Analog/Radio-Frequency Circuit Design Automation}}

\title{\Huge{RoSE-Opt: \underline{Ro}bu\underline{s}t and \underline{E}fficient Analog Circuit Parameter \underline{Opt}imization with Knowledge-infused Reinforcement Learning}}

% \title{\Huge{HERA: Highly Efficient and Reliable of \\Analog Circuit Parameter Optimization with Knowledge-Infused Deep Learning}}

% Optimization}}

% \underline{S}ampling-\underline{E}fficienct Domain 

\author{Weidong Cao,~\IEEEmembership{Member,~IEEE,}
Jian Gao,~\IEEEmembership{Student Member,~IEEE,} Tianrui Ma,~\IEEEmembership{Student Member,~IEEE,}\\
Rui Ma,~\IEEEmembership{Senior Member,~IEEE,}
Mouhacine Benosman,~\IEEEmembership{Senior Member,~IEEE,}
and Xuan Zhang,~\IEEEmembership{Senior Member,~IEEE}

% \thanks{Manuscript received July 30, 2022.
% }  
\thanks{This work was supported in part by
the National Science Foundation under Grant no. CCF-1942900.}
% \thanks{This work is done by W. Cao in his internship at Mitsubishi Electric Research Laboratories.
% }
\thanks{W. Cao is with the Department of Electrical \& Computer Engineering; The George Washington University, Washington, DC, 20052 USA, email: $\{$weidong.cao@gwu.edu$\}$. Part of this work is done by W. Cao in his internship with the Mitsubishi Electric Research Laboratories, Cambridge, MA, 02139 USA.}
\thanks{J. Gao and X. Zhang are with the College of Engineering; Northeastern University, Boston, MA, 02115 USA.}
\thanks{T. Ma is with the Department of Electrical and Systems Engineering; Washington University in St. Louis, St. Louis, MO, 63130 USA.} 
% \thanks{W. Cao, J. Gao, T. Ma, and X. Zhang are with the Department of Electrical and Systems Engineering; Washington University in St. Louis, St. Louis, MO, 63130 USA, e-mail: $\{$weidong.cao@wustl.edu, xuan.zhang@wustl.edu$\}$. Part of this work is done by W. Cao in his internship at Mitsubishi Electric Research Laboratories.} 
\thanks{M. Benosman is with the Mitsubishi Electric Research Laboratories, Cambridge, MA, 02139 USA.}
\thanks{R. Ma is with pSemi a Murata company, San Diego, CA, 92121 USA.}
\thanks{W. Cao and J. Gao equally contribute to the work.}

% 		Manuscript received May 01, 2021.
% 		%revised Month xx, xxxx; accepted Month x, xxxx.
% 		This work was supported in part by the xxx Department of xxx under Grant  (sponsor and financial support acknowledgment goes here) (Corresponding author: Rui Ma).
		
% 		W. Cao, and X. Zhang are with the Department of Electrical and Systems Engineering; Washington University in St. Louis, St. Louis, MO, 63130 USA, e-mail: $\{$weidong.cao@wustl.edu, xuan.zhang@wustl.edu$\}$.
		
% 		B. Mouhacine, and R. Ma are with the Mitsubishi Electric Research Laboratories, Cambridge, MA, 02139 USA, e-mail: $\{$benosman@merl.com, r.ma@merl.com$\}$.

}

\maketitle

\begin{abstract} 
Design automation of analog circuits has long been sought.
However, achieving robust and efficient analog design automation remains challenging.
This paper proposes a learning framework, RoSE-Opt, to achieve \underline{ro}bu\underline{s}t and \underline{e}fficient analog circuit parameter \underline{opt}imization.
RoSE-Opt has two important features.
First, it incorporates key domain knowledge of analog circuit design, such as circuit topology, couplings between circuit specifications, and variations of process, supply voltage, and temperature, into the learning loop.
This strategy facilitates the training of an artificial agent capable of achieving design goals by identifying device parameters that are optimal and robust.
{Second, it exploits a two-level optimization method, that is, integrating Bayesian optimization (BO) with reinforcement learning (RL) to improve sample efficiency.}
In particular, BO is used for a coarse yet quick search of an initial starting point for optimization.
This sets a solid foundation to efficiently train the RL agent with fewer samples.
{Experimental evaluations on benchmarking circuits show promising sample efficiency, extraordinary figure-of-merit in terms of design efficiency and design success rate, and Pareto optimality in circuit performance of our framework, compared to previous methods.}
Furthermore, this work thoroughly studies the performance of different RL optimization algorithms, such as Deep Deterministic Policy Gradients (DDPG) with an off-policy learning mechanism and Proximal Policy Optimization (PPO) with an on-policy learning mechanism.
This investigation provides users with guidance on choosing the appropriate RL algorithms to optimize the device parameters of analog circuits.
Finally, our study also demonstrates RoSE-Opt's promise in parasitic-aware device optimization for analog circuits.
{In summary, our work reports a knowledge-infused BO-RL design automation framework for reliable and efficient optimization of analog circuits' device parameters.}
Code implementation of our method can be found at \url{https://github.com/xz-group/RoSE}.

\end{abstract}

\iffalse
% Note that keywords are not normally used for peer review papers.

\fi

\IEEEpeerreviewmaketitle

\section{Introduction}
\label{sec:intro}

{I}{ntegrated} circuit (IC) technology advances human society by powering numerous applications and infrastructures with microelectronic chips of a small footprint.
Recent advances in deep learning have shown great promise in transforming modern IC design workflows~\cite{BO, google, nvidia, dongcktgnn}. 
By formulating each design stage as a learning problem, machine learning techniques can significantly shorten IC development cycles compared to conventional Electronic Design Automation (EDA) tools.
For example, Google~\cite{google} and Nvidia~\cite{nvidia} have shown that deep learning methods can improve design efficiency by an order of 100$\times$ at certain stages of the digital IC design flow, such as floor planning and power estimation.
Analog circuit is an essential type of circuit that bridges our physical world with the digital information realm~\cite{RFIC, IC, Cao2, Cao3, Cao5}.
Yet, unlike digital ICs that benefit from well-established conventional EDA tools or emerging efficient learning-based design automation methods, analog circuits continue to rely on onerous human efforts and lack effective EDA techniques at all stages~\cite{BO, GNN_distributed}.

{The pre-layout design of analog circuits can be represented as a parameter-to-specification (P2S) optimization problem.}
Given the topology of a circuit, the goal is to find optimal device parameters (e.g., width and finger number of transistors) to meet the desired specifications (e.g., power and bandwidth) of the circuit. 
This problem is challenging due to several factors.
First, it involves searching for parameters of diverse devices in a large design space.
The complexity grows exponentially with an increase in both design parameters and circuit specifications~\cite{RFIC, IC}. 
Second, the actual interactions between the device parameters and the circuit specifications are complicated~\cite{BO, GNN_distributed}, depending on multiple variables, such as circuit topology, variations in process, voltage, and temperature (PVT) and post-layout parasitic effects.
There are no exact analytical rules to follow, which worsens the search process.
Conventionally, human designers use critical domain knowledge, such as circuit topologies and couplings between circuit specifications, to manually derive the device parameters.
In particular, a human designer exerts an intense effort to obtain empirical equations between the device parameters and the circuit specifications based on a simplified circuit topology.
Yet, despite the simplification, tens and even hundreds of iterative fine-tunings are still required to ensure the accuracy and reliability of the design.

During the past several decades, there have been enormous explorations on automating the design of analog circuit device parameters.
These methods generally fall into two categories, knowledge-based techniques and optimization-based techniques. 
Knowledge-based techniques are designer-centric~\cite{BAG, IDAC, OASYS}.
They customize the design steps for specific circuits based on domain knowledge and embed them into procedural scripts that mimic the actions of designers.
These scripts allow designers to have full control over the modification and debugging of circuits to guarantee design reliability.
However, design efficiency is significantly thwarted, because designers, acting as optimization agents, are required to frequently interact with procedural scripts.
In contrast, optimization-based techniques are algorithm-centric.
They consider each step of analog circuit design as a black-box optimization problem and apply optimization methods, such as Bayesian optimization~\cite{BO}, Genetic algorithms~\cite{generic}, and emerging machine learning algorithms~\cite{GNN_distributed, supervised_1, sv1, sv2, RL_1, RL2}.
These algorithms can be run quickly to complete the design of an analog circuit with high efficiency.
Unfortunately, due to the absence of knowledge from experienced designers, the reliability of the design is not guaranteed, e.g., device parameters are not robust to various non-idealities.
These defects limit the efficiency and reliability of state-of-the-art analog design automation techniques.

To bridge this gap, we propose a learning framework, RoSE-Opt, to achieve \underline{ro}bu\underline{s}t and \underline{e}fficient analog circuit parameter \underline{opt}mization by synergizing domain knowledge of analog circuits and learning algorithms.
Analog circuit design strongly relies on domain knowledge, such as circuit topology, couplings between circuit specifications, and PVT variations; thus, without adequately considering these key domain knowledge in building learning-based design automation frameworks, the device parameters discovered by the algorithm are prone to suffer from inferior reliability issues due to various non-idealities.
Our previous work~\cite{rose} follows this principle and has explored the integration of this key domain knowledge into the design framework.
It also exploits a two-level optimization method by integrating Bayesian optimization (BO) and reinforcement {learning (RL) to improve sample efficiency.}

In this paper, we propose the RoSE-Opt framework which advances the state-of-the-art method~\cite{rose}.
In particular:
\circlednew{white}{1} We analyze the failed cases in which our trained RL agent cannot
converge to the optimal device parameters.
In these scenarios, the RL agent can still help designers by offering optimized initial points for manual tuning.
\circlednew{white}{2} We study its ability to consider device parasitic in parameter optimization.
A direct mapping of an analog circuit schematic with correctly-sized devices into a physical layout can lead to performance degradation, mainly due to parasitics from metal wires and electromagnetic effects, and needs tens of iterations between the two for fine-tuning.
This extended work demonstrates the promise of RoSE-Opt in addressing this problem.
\circlednew{white}{3} At the algorithm level, we thoroughly study the performance of different RL optimization algorithms, such as Deep Deterministic Policy Gradients (DDPG) with an off-policy learning mechanism and Proximal Policy Optimization (PPO) with an on-policy learning mechanism, to provide users with useful guidance in choosing appropriate RL algorithms for device sizing.
\circlednew{white}{4} {Finally, we showcase a Pareto optimization example, i.e., optimizing the figure-of-merit ($\text{FoM}$) of a circuit.
Our framework can facilitate the identification of the optimal Pareto frontier.
{In summary, we present a holistic BO-RL-based design automation framework to perform the P2S optimization tasks of analog circuit design with high robustness and efficiency.}
We make following key contributions.
\begin{itemize}
%\vskip -0.25cm

\item This paper proposes a comprehensive BO-RL-based design automation framework, RoSE-Opt.
Our learning framework explores and exploits both the domain knowledge of analog circuit design and the strong optimization ability of design automation algorithms.
 
\item We perform a failure analysis of our method and show how to leverage the unsuccessful deployment trajectory to guide the fine-tuning of manual efforts toward design success.
In addition, we study its effectiveness in the scenarios of parasitic-aware device parameter optimization.

\item We thoroughly characterize the performance of different RL optimization algorithms (i.e., DDPG vs. PPO) to provide users with useful insights in choosing appropriate RL algorithms for device parameter optimization.

\item Experimental evaluations on benchmarking circuits show that our framework achieves $7.9\times\sim12\times$ improvement in training sample efficiency and a significant improvement in the design success rate and efficiency compared to the state-of-the-art methods for the same problem. 
In Pareto optimization, our framework identifies the optimal Pareto frontier by minimizing power consumption and maximizing the gain bandwidth product (GBW), all within a minimal number of simulations. 

\end{itemize}

\section{Background and Related Work  }
\label{sec: re_work}

In this section, we first review the basics of Bayesian optimization and reinforcement learning.
{We then introduce the key domain knowledge that human experts commonly consider when addressing the P2S problem.}
Finally, we discuss existing design automation methodologies for analog circuits.

\subsection{Bayesian Optimization}

Bayesian optimization (BO) proves to be a valuable framework to address challenging black-box optimization problems that involve costly function evaluations. 
Fig.~\ref{fig: rl_intro}(a) shows an example of BO with two iterations ($t=2$ and $t=3$).
BO's fundamental concept is to construct an inexpensive surrogate model, such as a Gaussian Process, by leveraging actual experimental data. 
This surrogate model incorporates prior knowledge or beliefs about the objective function, which is then used to make informed decisions in the process of selecting a sequence of function evaluations through the use of an acquisition function, such as expected improvement (EI).
It also balances exploration and exploitation. 
Exploration allows for a broader exploration of the search space, potentially discovering better solutions, while exploitation focuses on exploiting the known promising areas to optimize the current best solution. 
Balancing these two aspects is crucial to finding better solutions and refining the best solution.

Given an arbitrary function $f(\vec{x})$ for maximization, there are several steps to follow for BO.
\noindent{{Step 1: initial sampling}.} Here, a limited set of sample points is randomly selected.
\noindent{{Step 2: initializing the model}.} These points in Step 1 are used to calculate a surrogate function.
\noindent{{Step 3: iterating}}. In particular, the acquisition function is first used to get the next point; then, the surrogate function is re-evaluated; third, the surrogate function is verified to see if it remains stable or if the variance falls below a predetermined threshold, or if $f(\cdot)$ is exhausted, depending on the specific design objective.

BO is well suited to optimizing hyperparameters of many classification and regression models. \textcolor{black}{It is also used for circuit design automation~\cite{BO,ghaffari2023statistical,bai2021boom,liu2019accelerating} (e.g., the P2S optimization tasks for analog circuit design~\cite{BO}).}

\begin{figure}[!t]
\begin{center}
\includegraphics[width=0.78\linewidth]{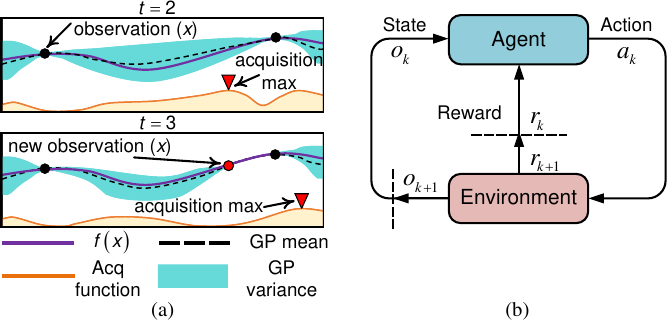}
\vskip -6pt
\caption{(a) An illustration of Bayesian optimization to find the optima. Here, we use the Gaussian Process (GP) as the surrogate model and show two iterations. The plots show the mean and confidence intervals estimated with the GP model of the objective function, $f(x)$, which in practice is unknown. The plots also show the acquisition (Acq) functions in the lower-shaded plots. The acquisition is high where the model predicts a high objective (exploitation) and where the prediction uncertainty is high (exploration). (b) A simplified illustration of reinforcement learning. It includes five parts: agent, action, state, reward, and environment.}
\label{fig: rl_intro}
\end{center}
\vskip -18pt
\end{figure}

\subsection{Reinforcement Learning}
\label{sec: rl}
Reinforcement learning (RL) is a machine learning method related to how intelligent agents take actions in an environment to maximize cumulative returns based on states.
As illustrated in Fig.~\ref{fig: rl_intro}(b), there are five essential elements in an RL problem: Agent, Action, State, Reward, and Environment.
The `Agent' is the learner and the decision maker who learns experiences from a training process and makes decisions based on observations (states) from the environment.
The `Action' is a set of operations that the agent can perform.
%in a state.
The `State' is a representation of the current environment (i.e., observations) in which the agent is staying.
It can be observed by the agent and contains all relevant information about the environment that the agent needs to know to make a decision.
The `Reward' is a scalar value returned by the environment after the agent takes an action in a state.
It is used to evaluate and guide the actual learning behavior of the agent.
The `Environment' is the physical world in which the agent operates.

In each episode, an agent starts from an initial state, then observes the state $o_k$ and takes an action $a_k$ based on a policy. 
Meanwhile, the environment updates a reward $r_{k+1}$ for that particular action and enters a new state $o_{k+1}$.
The agent iterates through the episode in multiple steps, accumulating the reward at each step to obtain the final return.
With multiple episodes, the RL agent improves its decision quality and finds the best policy to maximize the return.
Such a policy would be deployed for practical tasks, i.e., the agent follows the trained policy to finish a given task.

RL algorithms have been extensively applied to many problems such as game playing~\cite{game}, robotics~\cite{robot}, computer vision~\cite{CV}, and natural language processing~\cite{dl_nlp}.
{RL has also been used to automate the design of ICs, such as the placement of the digital IC chip~\cite{google_chip_placement}, and the P2S optimization of analog circuits~\cite{RL1, RL_1, RL2}.}

\subsection{Key Domain Knowledge of Analog Circuit Design}
\label{sec: dk}

At the pre-layout stage, there are many considerations to be taken by human experts to select reliable device parameters and meet the design goals.
These considerations are the domain knowledge, and we introduce the major ones that are commonly used by human experts when they tackle the P2S optimization tasks, as shown in Fig.~\ref{fig: human_loop}.
\subsubsection{Circuit topology} When human experts manually find the optimal device parameters, they first construct the circuit small-signal model from the circuit topology, based on which they obtain empirical equations that connect the device parameters to the circuit specifications.
With these equations, device parameters can be derived by hand.
\subsubsection{Couplings between circuit specifications} Due to design trade-offs, circuit specifications often depend on each other. 
For example, in the design of operational amplifiers, energy efficiency often trades off with gain; that is, a higher amplification gain requires a larger transconductance, which, however, demands more power consumption and results in lower energy efficiency.
Therefore, in a conventional manual design process, human experts use tens and even hundreds of iterative fine-tunings to find a group of proper device parameters to satisfy all circuit specifications. 
\subsubsection{PVT variations}
To ensure the robustness of analog circuits in different harsh environments, a key design consideration is to minimize the influence of variations in process (P), voltage (V), and temperature (T).
Process variation represents the deviation of the manufactured devices from their ideal specification due to manufacturing errors.
It includes \underline{t}ypical N-type transistor/\underline{t}ypical P-type transistor (TT), \underline{f}ast N-type transistor/\underline{f}ast P-type transistor (FF), \underline{s}low N-type transistor/\underline{s}low P-type transistor (SS), \underline{s}low N-type transistor/\underline{f}ast P-type transistor (SF), and \underline{f}ast N-type transistor/\underline{s}low P-type transistor (FS).
Voltage and temperature variations are due to uncertain ambient changes.
Typical deviation of the supply voltage is $\pm 10\% $ from its nominal value $V_{\text{DD}}$;
and typical range of the environmental temperature for circuits is $[-40, 125] ^\circ C$.
A single PVT corner is a combination of P, V, and T from their varying ranges.
All these variations are unavoidable and can cause the circuit performance degeneration compared to its nominal case, i.e., \{TT, $V_{\text{DD}}$, $25 ^\circ C$\}.
Manual experts have to look for robust device parameters to achieve the design goal in all PVT corners.

\subsubsection{Parasitic effects of physical layouts}
A complete flow of analog circuit design includes the schematic design and the physical layout design.
The conversion of an analog circuit schematic, with the correctly-sized components, into a physical layout can cause performance degradation due to the parasitic effects of metal wires and electromagnetic couplings.
Experienced human designers often make efforts to adjust the device parameters to ensure that the post-layout simulation meets the desired objectives.

\begin{figure}[!t]
\begin{center}
\includegraphics[width=0.78\linewidth]{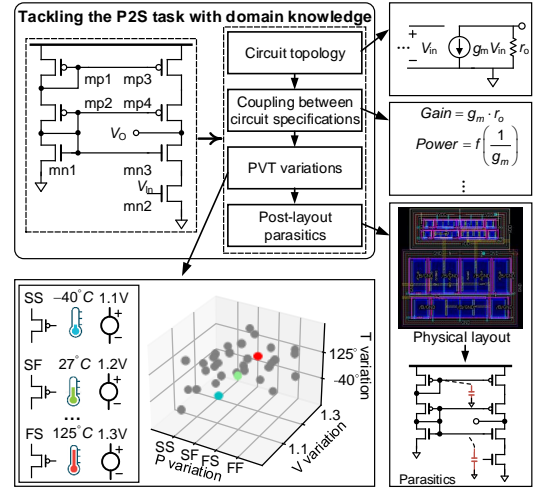}
\vskip -6pt
\caption{{Illustration of a manual design flow to tackle the P2S optimization tasks with human domain knowledge.}
}
\label{fig: human_loop}
\end{center}
\vskip -15pt
\end{figure}

\subsection{Existing Design Automation Methodologies}
\label{sec: ex_aeda}

{Various design automation techniques have been proposed for the P2S tasks of analog circuits in recent years.}
They generally fall into two categories: knowledge-based techniques and optimization-based techniques.
Knowledge-based techniques, such as BAG~\cite{BAG}, are designer-centric.
They tailor the design steps for specific circuits with domain knowledge and embed these steps into the procedural scripts that mimic designer actions.
These scripts provide designers with complete control over circuit modifications and debugging to ensure design reliability. 
Yet, these approaches notably affect design efficiency, as they demand frequent interactions between designers and procedural scripts, with designers playing the role of optimization agents.
On the contrary, optimization-based methods such as BO~\cite{BO}, Geometric Programming~\cite{geo_programming}, Genetic algorithms~\cite{generic}, and modern machine learning approaches~\cite{GNN_distributed, supervised_1, sv1, sv2, RL1, RL_1, RL2} are centered on algorithms. 
They treat each step in a circuit design as a black-box optimization problem and can swiftly perform optimization procedures to complete a circuit's design with high efficiency.
Unfortunately, the lack of knowledge from seasoned designers means that design reliability, such as the robustness of device parameters to non-ideal conditions, is not assured.
These limitations significantly impact the widespread applications of state-of-the-art analog design automation techniques, as they are unable to achieve both high design efficiency and reliability.
\textcolor{black}{Thus, essential to advance analog design automation is to adequately incorporate design knowledge into optimization algorithms to ensure design reliability while maintaining high optimization efficiency.}

Learning-based optimization methods have recently emerged. 
They show higher design efficiency and \textcolor{black}{scalability} in handling the P2S task compared to classical optimization algorithms such as BO~\cite{BO}, Geometric Programming~\cite{geo_programming}, and Genetic algorithms~\cite{generic} \textcolor{black}{that are limited to ad-hoc tasks and tackling simple circuits}. 
As an example, supervised learning methods~\cite{GNN_distributed,supervised_1, sv1,sv2} have been used to learn the complicated relations between device parameters and circuit specifications.
Once trained, they adopt one-step inference to predict optimal device parameters for given design goals. 
Nonetheless, these supervised learning methods cannot guarantee a high design success rate and suffer from weak generalization abilities~\cite{GNN_distributed,supervised_1, sv1,sv2} due to their inherent approximation errors.
On the other hand, RL methods~\cite{RL1, RL_1, RL2} learn an optimal policy from the state space of circuit specifications to the action space of device parameters, which solves a quasi-dynamic programming problem.
They often use multiple sequential decision steps to find the optimal device parameters rather than just using one-step prediction, thus achieving a higher design success rate and better generalization abilities than supervised learning methods~\cite{GNN_distributed,supervised_1, sv1,sv2}. 
However, none of them has taken into account sufficient domain knowledge of analog circuit design in the optimization loop, leading to low design reliability.

In this work, \textcolor{black}{we propose a learning-based framework, RoSE-Opt, to achieve efficient and reliable parameter optimization of analog circuit devices by harnessing the synergy between the knowledge of human designers and RL algorithms (elaborated in Section~\ref{sec:RL}).
In particular, we leverage the rapid convergence of BO to identify an optimized starting point, significantly improving the sampling efficiency of the primary RL agent during its learning phase.}

\begin{figure*}[!t]
\begin{center}
 \vskip -12pt
\includegraphics[width=0.8\linewidth]{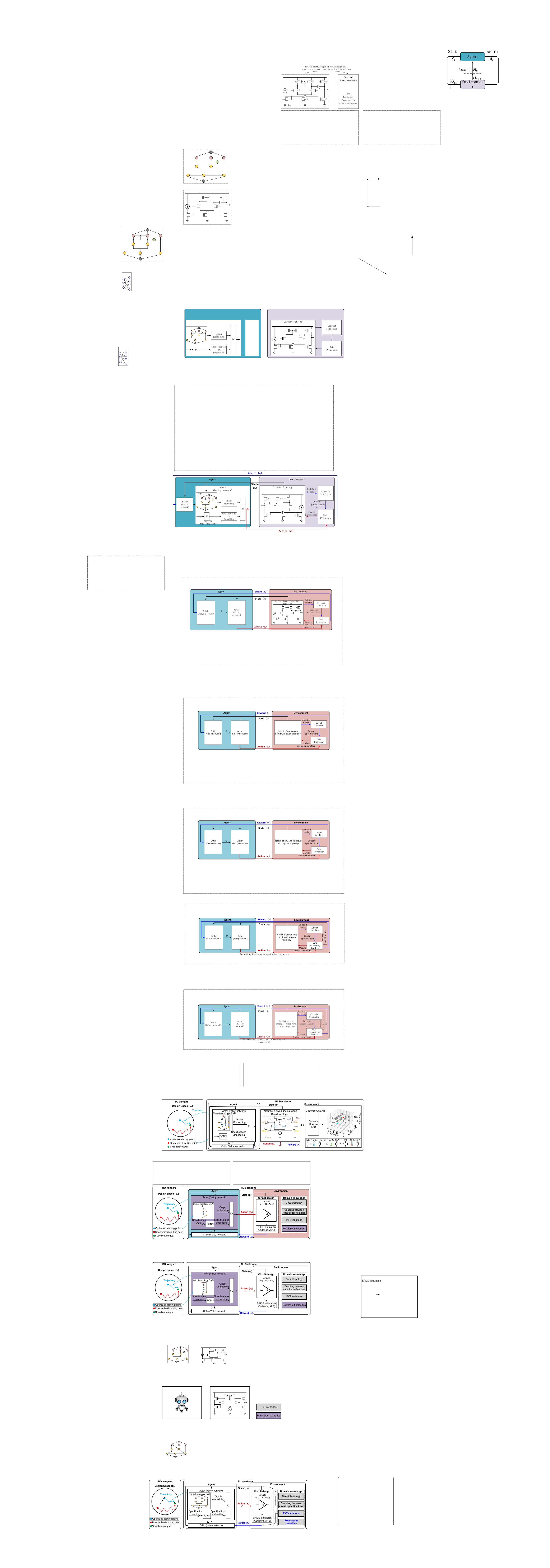}
\vskip -3pt
\caption{Overview of our RoSE-Opt framework for automated design of analog circuits by complementing BO vanguard and RL backbone. At its core, the framework leverages BO's rapid convergence to identify an optimized starting point for RL, significantly enhancing its sample efficiency throughout the learning or optimization process. This strategy combines the efficient exploration capabilities of BO with the robust optimization power of RL to ensure both design robustness and efficiency. The RL backbone is based on an actor-critic method. The environment consists of a netlist of an analog circuit with a given topology, a circuit simulator, and a data processor. At each time step $k$, the agent automatically produces an action $a_k$ to update device parameters with its policy network according to the state $o_k$ and then receives the reward $r_k$ from the environment. Our customized policy network is composed of a circuit topology-based GNN (i.e., GAT) and an FCNN.}
\label{fig: overview}
\end{center}
\vskip -15pt
\end{figure*}

\section{RoSE-Opt Framework}
\label{sec:RL}

{In this section, we introduce the proposed RoSE-Opt framework that automates P2S tasks.}
We start with the formulation of the problem.
Then, an overview of the RoSE-Opt framework is presented, followed by an elaboration of the BO vanguard.
Finally, we introduce five essential parts of the RL backbone and show how key domain knowledge is incorporated into the framework.

\subsection{Problem Formulation}

We target the P2S problem with a given circuit under stringent PVT variations and parasitic effects of physical layout, formulated as
\begin{equation}
\begin{aligned}
\min_{s} \quad & f(s,g),\\
\textrm{s.t.} \quad & s = F(x), ~
\text{where}   & s \in \mathbb{R}^{i \times j},
~x \in S_{P},  
 ~g \in S_{G}.
\end{aligned}
\label{eq: q_formulation}
\end{equation}
Here, the function $f(s,g)$ represents the difference between the circuit specifications $s$ and the design goal $g$. 
For example, for operational amplifiers (Op-Amps), there are four main circuit specifications, i.e., gain (G), power consumption (P), phase margin (PM), and bandwidth (BW).
$F(\cdot)$ is a circuit simulator environment to obtain circuit specifications $s$ based on a set of device parameters $x$, e.g., width and finger numbers of transistors.
$s$ is essentially a matrix where $i$ represents the specification type (e.g., gain) and $j$ represents a PVT corner. 
Therefore, $s \in \mathbb{R}^{4 \times 16}$, assuming $16$ PVT corners for the design of Op-Amps.
The set of device parameters $x$ is restricted by the design space $S_{P}$.
The design goal $g$ is restricted to a reasonable sampling space $S_{G}$ that the circuit can achieve.
Our objective is to minimize $f(s,g)$ by efficiently looking for a group of optimal device parameters so that the circuit specifications can meet an arbitrarily given group of design goals under all PVT variations.
Considerations of parasitic effects are discussed in Section~\ref{sec: para_sizing} as it is considered during the deployment stage rather than the training stage.

\subsection{Framework Overview}
\label{sec:RL_framework}

We explore the synergy of BO and RL to achieve robust and sampling-efficient device parameter optimization.
Fig.~\ref{fig: overview} shows the overview of the proposed RoSE-Opt framework, which contains two parts: a BO Vanguard and an RL Backbone.
BO is a well-known optimization algorithm that often achieves the fastest convergence~\cite{BO} to an optimum (or sub-optimum) for a given design goal, compared to other optimization techniques~\cite{geometric, generic}.
However, it needs to be restarted from scratch if the given design goal is changed and is often limited to tackling simple circuits with fewer dimensions.

In contrast, well-trained RL agents can reach general design goals without retraining based on a deployment trajectory from a starting point.
Unfortunately, for robust analog circuit design, which is a more complex problem, RL methods demand more data points from time-consuming circuit-level simulations (i.e., PVT simulations) to sufficiently explore design space, leading to a low sampling efficiency toward the convergence. With this key insight in mind, we propose to leverage BO as a vanguard to first search coarsely for a suboptimal starting point (i.e., initial device parameters) for our RL agent. On this basis, the RL agent can be trained to find optimal solutions with much fewer interactions with time-consuming circuit-level simulations, improving the sampling efficiency. As conceptually shown in the left subset of Fig.~\ref{fig: overview} (i.e., BO vanguard), an optimized starting point can help the RL agent reach design goals with a shorter trajectory compared to a randomly selected one.
Hence, it can guide the RL agent to converge faster with fewer training data.

The RL backbone has five essential components similar to typical RL methods (refer to Section~\ref{sec: rl}): reward, action space, state space, environment, and agent.
To train an excellent RL agent for a given task, there are several critical factors to pay attention to.
The first is to develop a comprehensive environment that could expose environmental information about the task to the RL agent as much as possible.
The second is to capture sufficient exposed observations (states) relevant to the task from the environment into the learning loop.
The third is to design a proper reward function that is closely related to the optimization goal and stimulates the learning of the RL agent. Finally, the agent (i.e., policy network) itself should be an expressive and powerful model that can distill the underlying domain knowledge and generalize it to unseen design goals. With these key factors in mind, we briefly introduce here (i) how to develop the majority of these five components below and (ii) how to distribute the domain knowledge of analog circuit design presented in Section~\ref{sec: dk} to the reward function and the customized policy network.
More details will be discussed in Section~\ref{sec: rl_backbone}.

\noindent\textbf{Comprehensive Environment.}
First of all, we develop a thorough circuit design environment for the P2S task, which includes the full circuit netlist of the given task and commercial simulation/verification tools (e.g., Cadence Spectre) for simulating circuit specifications (under PVT variations) and extracting post-layout parasitics.

\noindent\textbf{Sufficient Observations.}
Second, the circuit design is a dynamic process that needs sufficient observation from its simulation environment. In particular, for a robust analog design, full PVT corner simulations are essential to ensure training stability. Thus, our RL's state consists of dynamic intermediate circuit specifications from all PVT corners and the corresponding circuit topology/device parameters. 

\noindent\textbf{Custom Reward Function.}
Third, PVT variations and parasitic of post-layouts affect the circuit specifications, which are directly related to the optimization goal. We thus use a custom reward function to take into account PVT variations (refer to Eq.~\eqref{eq: reward} for details). 
By infusing key domain knowledge (e.g., PVT variations) into the RoSE-Opt framework through the reward function, an excellent RL agent can be trained and make good decisions to search for reliable device parameters that meet the design goals.

\noindent\textbf{Customized expressive policy network model.}
Lastly, we customize a policy model architecture to enhance its expressiveness by taking in sufficient observations. 
In particular, we tailor a novel circuit topology-based GNN and an FCNN to incorporate into the learning loop of an RL agent.
This policy network can effectively capture the essential physical features (e.g., device parameters and interactions) embedded in a circuit graph with the GNN and extract the couplings (i.e., design trade-offs) between circuit specifications with the FCNN, which better models the relations between the circuit parameters and the design targets.

During the training of an RL agent, in each episode, the agent starts from an initial state $o_0$ with a group of initial device parameters optimized by the BO vanguard and a group of randomly-sampled desired specifications $g$ from sampling space $S_{G}$.
The end of an episode occurs when the design goals are achieved or a predefined maximum step $T$ is reached. 
At each time step $k$, the agent begins using a neural network to observe a state $o_k$ and take discrete action $a_k$ based on the probability distribution of the output of the neural network. 
The agent then arrives in a new state $o_{k+1}$ and receives a reward $r_k$ from the environment. 
The discrete action $a_k$ can update simultaneously all the device parameters for the given circuit.
The agent iterates through the episode with multiple steps and accumulates the reward at each step until the end of the episode. 
In the next episode, the agent randomly samples another design goal $g$ from the sampling space $S_{G}$ and resets the parameter back to the starting point $o_0$. 
Then repeat the same process again. 
Once the policy network is well trained, we can save the weight of the neural network for deployment. 
During the deployment, since the weight has already been trained, the agent uses only the actor to take actions based on the state it observed. 
The purpose of the deployment part is to show the generalization capability of our trained policy network to different specifications without retraining like BO. 
Thus, we are interested to see how many specifications the decision policy can reach within the predefined maximum step $T$ and what is the average deployment length for each run.

A key point is that BO is only required once in our framework if the sampling space $S_{G}$ of the design goals and the design space $S_{P}$ of each circuit device are defined.
The RL agent then uses the same optimized starting point $o_0$ during all training episodes and the deployment stage.
Note that in the context of robust device sizing, the designs of both the BO vanguard and the RL backbone are non-trivial and are elaborated in the following.

\subsection{BO Vanguard}
\label{sec: label}
We rely on BO to find an optimized initial search point for our RL agent to improve its sampling efficiency during training.
However, a crucial initial question is how to define such an optimized starting point. 
This starting point should not only speed up the design for a specific set of goals but should also help, in general, to efficiently design any arbitrary group of design goals from the entire sampling space $S_P$.

To solve the problem, we think that from this starting point, the RL agent should generally take the least deployment steps to achieve a general design.
Thus, we let the device parameters found by BO that achieve as closely as possible the arithmetic mean of the maximum/minimum of each design goal in the entire sampling space $S_G$, i.e., $({PM_{\max} + {PM}_{\min}})/2$, $({{G}_{\max} + {G}_{\min}})/2$, $({BW}_{\max} + {BW}_{\min} )/2$, and $({P}_{\max} + {P}_{\min} )/2$, be the starting point of our RL agent.
Here, taking a two-stage Op-Amp as an example, $G,~B,~PM,~P$ are the circuit specifications, i.e., gain ($G$), bandwidth ($B$), phase margin ($PM$), and power consumption ($P$).
Another important question is how to balance the simulation budget between BO and RL. 
Our strategy is to monitor the improvement of the reward function every 10 iterations and stop BO optimization if the improvement falls below a specified threshold (i.e., 0.002).
%\footnote{\textcolor{blue}{This threshold is a hyperparameter to balance the simulation budget between BO and RL. In experiments, we often set this threshold to 0.002.}}.
Users can also adjust the maximum number of simulations assigned to BO, typically set to 50, to ensure efficient resource management. BO is stopped when either is reached. 
Both hyperparameter values are set based on our empirical study.

We use a typical set-up of BO to search for the optimized starting point, which includes two essential parts: the surrogate model and the acquisition function~\cite{BO}.
The whole optimization depends on how accurately the surrogate model estimates the black-box function. 
In particular, we adopt the widely used Gaussian process (GP) model as our surrogate model to predict the underlying black function with uncertainty.
We use a Monte-Carlo-based Expected Improvement (EI) acquisition function to balance exploration and exploitation during the optimization by offering the next sampling point as below:
\begin{equation}
\begin{aligned}
\quad & \mathrm{EI}(X) \approx \frac{1}{Z} \sum_{i=1}^Z \max _{j=1, \ldots, n}\left\{\max \left(\xi_{i j}-f(s,g)_{\text{best}}, 0\right)\right\}, \\
& \quad \xi_i \sim \mathbb{P}(f(X) \mid \mathcal{D}).
\end{aligned}
\label{eq: acfunction}
\end{equation}
Here, the expectation, $\text{EI}(X)$, is computed by approximating the integrals over the posterior distribution over $Z$ points using Monte-Carlo sampling. $\mathbb{P}(f(X)\mid\mathcal{D})$ is the posterior distribution of our function $f(s,g)$ at $X$ where $X=\left(x_1, \ldots, x_n\right)$ from the sampling in our design space $S_{P}$. $\mathcal{D}$ is our data set. 
The parameter $\xi$ determines the amount of exploration during optimization. 
% \textcolor{blue}{Using the Monte-Carlo-based EI acquisition function helps us to approximate the integral over the observed function's posterior distribution for high dimensional problems.}

\subsection{RL Backbone}
\label{sec: rl_backbone}

The five key components of the RL backbone are detailed below.

\subsubsection{Variation-aware reward function}
\label{sec: reward}
We connect the objective in Eq.~\eqref{eq: q_formulation} to our reward function so that our RL agent can be directly optimized considering PVT variations.
Particularly, the reward $r_k$ at each time step $k$ is designed by taking PVT variations into consideration, i.e.,
\begin{equation}
\begin{aligned}
\quad & r_k = \text{Mean} \big( \sum_{j=0}^{j=M-1} r_j  \big); ~\text{if~} \exists j\in[0,~ M-1], ~r_j<0; \\
& \text{or~} r_k=R, ~\text{if~}\forall j \in [0,~ M-1], ~r_j =0.
\end{aligned}
\label{eq: reward}
\end{equation}
Here, $r_j=\sum_{i=0}^{N-1} w_{i} \times  \min \{ {(s_i^j-g_i)}/{(s_i^j+g_i)}, 0\}$ is the sub-reward of the $j^{\text{th}}$ corner, calculated based on a weighted sum of the normalized difference between $i^{\text{th}}$ intermediate
circuit specification of the $j^{\text{th}}$ corner $s_i^j$ and $i^{\text{th}}$ design goal $g_{i}$.
All types of circuit specifications are equally important, i.e., $w_{i} = 1$.
$M$ represents the number of PVT corners and $N$ indicates the number of circuit specifications.
In order not to over-optimize the specification, we set the upper bound of $r_j$ to be $0$.
Only when the circuit specifications in all PVT corners meet the design goal, a large stimulated reward of $R = 10$ is given to encourage the agent for the successful design;
otherwise, the reward in each time step is the average of sub-rewards of all PVT corners.
Finally, the accumulative reward for a training episode is $R_{s,g} = \sum_{k=1}^{T}r_{k}$, where $T$ is a pre-defined maximum step for an episode.
Intermediate circuit specifications matrix $s$ is obtained from our high-fidelity simulation environment $F(\cdot)$ based on the updated device parameters $x$ at each time step.
Therefore, our reward is a direct measurement from the circuit simulator, which can help train a high-quality RL policy network.

\subsubsection{Fine-grained action}
\label{sec: action}
Inspired by human designers who rely on multiple fine-grained tuning steps to find optimal device parameters, we use a discrete action space to tune device parameters.
For each tunable parameter $x$ of a device (e.g., the width and finger number of transistors, the capacitance of capacitors), there are three possible actions: increasing ($x+\triangle x$), keeping ($x+0$), or decreasing ($x-\triangle x$) the parameter, where ``$\triangle x$" is the smallest unit used to update the parameter within its bound $x\in[x_{\min},~x_{\max}]$.
With the total parameters of $M$ devices\footnote{We often use differential pairs to reduce the number of design variables.}, the output of the policy network is a matrix of probability distribution $M\times 3$ in any state where each row corresponds to a parameter.
The action is taken based on the probability distribution.

\subsubsection{Circuit physics-related state}
\label{sec: state}

RL belongs to representation learning.
Capturing adequate state information from the environment is key to training an excellent RL agent. 
We leverage the domain knowledge of analog circuits, i.e., intermediate circuit specifications and circuit topology, as our state, which covers the most essential observations from a circuit design environment.
In particular, we take care of intermediate circuit specifications in all PVT corners, in contrast to the previous work~\cite{RA} which only considers partial PVT corners.
We create a state vector to represent the intermediate circuit specifications.
%, which is used as input for the FCNN part in our policy network. 
To better use the observations of the circuit itself, we use a graph $G(V,~E)$ to model the circuit according to its topology, where each node in the set $V$ is a device and the connections between the devices constitute the edge set $E$.

\begin{figure}[!t]

\includegraphics[width=1.0\columnwidth]{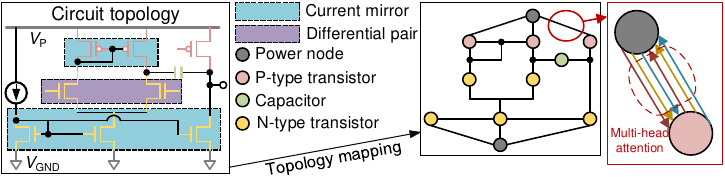}
\vskip -3pt
\caption{
Mapping circuit topologies to graphs and illustrating the tailored GAT in the policy network for analog circuit design, using a two-stage Op-Amp as an example.
}
\label{fig: policy_network}
%\end{center}
\vskip -15pt
\end{figure}

Fig.~\ref{fig: policy_network} shows the mapping between the circuit topology and a graph taking a two-stage Op-Amp as an example.
For a circuit with $n$ nodes, the state of the $i^{\text{th}}$ node is defined as its node feature $(t,~\vec{p})$, where $t$ is the binary representation of the type of node and $\vec{p}$ is the parameter vector of the node.
Note that the parameters of the circuit device reflect the physical information of the circuit.
For transistors, the parameters are the width ($x_{\text{W}}$) and the number of fingers ($x_{\text{F}}$).
For capacitors, resistors, or inductors, the parameters are scalar values (e.g., capacitance, resistance, or inductance) of each device.
For example, for a circuit with five different types of devices, the state of a $N$ type transistor can be expressed as $[{0,~0,~1},~ {x_{\text{W}},~x_{\text{F}}}]$.

\subsubsection{SPICE simulation environment}
\label{sec: env}

In our work, a high-fidelity circuit design environment with PVT variations and post-layout parasitics is used. 
It consists of the netlist of a given analog circuit, a commercial circuit simulator, e.g., Cadence Spectre for CMOS analog circuits or Keysight Advanced Design System (ADS) for RF power circuits, and a data processing module (DPM).
As shown in Fig.~\ref{fig: overview}, the simulator obtains intermediate circuit specifications at each time step.
The DPM then deals with the simulated results to return a reward to the agent using Eq.~\eqref{eq: reward}.
Meanwhile, it also updates the device parameters to rewrite the circuit netlist based on the actions of the agent (i.e., policy network).

Previous methods~\cite{RA} assume that the circuit simulation time scales linearly with the number of simulations, i.e., the simulation time for $16$ PVT corners is $16\times$ of the one with a single PVT corner.
We use the Cadence Spectre Accelerated Parallel Simulator (APS) to accelerate our simulation.
At each time step, we obtain circuit specifications for all PVT corners.
Compared to a single PVT corner time, the batch simulation manner for $16$ PVT corners only brings $0.17\times$ time overhead as compared to a single PVT simulation.
In other words, our circuit environment can achieve a sampling efficiency of at most $14\times$ when collecting data points during training compared to previous RL methods.
With this co-design loop, we are able to simultaneously achieve both high sample efficiency and robust design by taking advantage of BO, RL, and the simulation environment.

\subsubsection{Circuit-aware policy network}
\label{sec:agent}

We adopt an Actor-Critic method~\cite{actor_critic} to design our agent. 
To capture sufficient observations from the environment to the learning loop, we tailor a novel multimodal policy network for the Actor by integrating several unique features, as shown in Fig.~\ref{fig: overview}. 
(i) We customize a new network architecture consisting of two different networks, a GNN based on the circuit topology and an FCNN.
It is termed a GNN-FC-based policy network. 
Specifically, the GNN is used to distill the underlying physics (e.g., device types, parameters, and interactions) of a circuit graph into a low-dimensional vector embedding. 
The FCNN utilizes the design objectives and intermediate circuit specifications across all PVT corners as input to reveal their interconnected relationships, i.e., design trade-offs. 
The graph embedding and the FCNN embedding are then concatenated for further processing by the final FC layers to update the actions.
The value network (Critic) has the same architecture as the policy network except for the last layer. 
It is used to evaluate the quality of the actor's decision by giving an estimate of the expected reward, $Q$, for the execution of the current policy. 
(ii) We use dynamic device parameters to encode node features.
Yet, the node features of the previous GNN-based method~\cite{RL2} are composed solely of static technological data, such as threshold voltage and electron mobility. 
The exclusion of crucial dynamic device parameters from the node features complicates the process of learning the relationships between device parameters and design goals and leads to divergence of learning in our study.
(iii) We choose the graph attention network (GAT)~\cite{gan} as the backbone of the GNN part, whereas previous work~\cite{RL2} often applies the graph convolutional network (GCN) to solve the problem.
The multi-head attention mechanism of GAT helps to learn more complex and higher-dimensional interactions between a circuit node and its neighbors (refer to Fig.~\ref{fig: policy_network}).
Our empirical studies (elaborated in Section~\ref{sec:exp}) show that GAT often performs better than other GNNs such as the GCN~\cite{gcn} in the P2S task.
Note that we customize the GAT to model circuit topologies and apply them to the P2S task rather than inventing novel GNN structures. 
The fundamental operations underlying the proposed GAT follow those of the original publications~\cite{gan} and, therefore, are omitted here. With this unique policy network architecture, each circuit has its specific state representation: a vector consists of its design specifications (used as the input for the FCNN part) and a matrix with the circuit node feature encoding (used as the input for the GNN part).
We perform thorough experimental studies in Section~\ref{sec: robustness and sample efficiency} to show that the customized network architecture is superior to other methods.

\begin{table*}[ht]
  \centering
  %\vskip -0.6in
\captionof{table}{Design space, sampling space, and PVT corners for four benchmark circuits. 
}
\vskip -3pt
% \vskip -0.05in
\begin{footnotesize}
\begin{tabular}{c|cccc|cccc|cccc|cccc}

\toprule
Circuit types & \multicolumn{4}{c|}{Single-stage Op-Amp} & \multicolumn{4}{c|}{Two-stage Op-Amp} & \multicolumn{4}{c|}{Folded-cascode Op-Amp}  & \multicolumn{4}{c}{Nested Miller compensation Op-Amp}\\
\hline
Technology   &\multicolumn{16}{c}{GlobalFoundries $130/65/28$ nm} \\
\hline
16 PVT conditions   &\multicolumn{6}{c}{Process:\{SS, SF, FS, FF\}} 
&\multicolumn{6}{c}{Voltage:\{$1.1$V, $1.3$V\}}
&\multicolumn{4}{c}{Temperature:\{$-40^\circ C$, $125^\circ C$\}} \\
\hline
%\# of devices & \multicolumn{4}{c|}{$8$} & \multicolumn{4}{c|}{$12$}  & \multicolumn{4}{c}{$17$}\\
%\hline
Design space &\multicolumn{4}{c|}{$10^{24}$ values}&\multicolumn{4}{c|}{$10^{31}$ values}&\multicolumn{4}{c|}{$10^{17}$ values}&\multicolumn{4}{c}{$10^{39}$ values}\\
Width (nm)&\multicolumn{4}{c|}{mp1-4:[200, 2000, 10]}&\multicolumn{4}{c|}{mp1-2:[1000, 100000, 1000]}  &   \multicolumn{4}{c|}{mp1-2:[1000, 10000, 200]}  &   \multicolumn{4}{c}{mp1-3\&mn4:[10000, 50000, 1000]}\\
&\multicolumn{4}{c|}{mn1-3:[2000, 10000, 10]}&\multicolumn{4}{c|}{mn1-4: [1000, 100000, 1000]}  &   \multicolumn{4}{c|}{mn1-3:[160, 1000, 20]} &   \multicolumn{4}{c}{mp4: [50000, 250000, 10000]} \\
&\multicolumn{4}{c|}{}&\multicolumn{4}{c|}{}  &   \multicolumn{4}{c|}{}  &   \multicolumn{4}{c}{mn1-3:[2000, 20000, 1000]}\\
Capacitance (pF)&\multicolumn{4}{c|}{$C_{L}$: 0.12}&\multicolumn{4}{c|}{c:[0.1, 10, 0.1] / $C_{L}$: 1}  & \multicolumn{4}{c|}{c:[0.1, 10.0, 0.2] / $C_{L}$: 1}& \multicolumn{4}{c}{c1:[25, 50, .5]/c2:[1, 25, .5]/$C_{L}$: 100}\\
% &\multicolumn{4}{c|}{}&\multicolumn{4}{c|}{$C_{L}$: 1}  &\multicolumn{4}{c|}{$C_{L}$: 1} &\multicolumn{2}{c}{c2:[1, 25, .5]}&\multicolumn{2}{c}{$C_{L}$: 100}\\
% &\multicolumn{4}{c|}{}&\multicolumn{4}{c|}{}  &\multicolumn{4}{c|}{} &\multicolumn{4}{c}{$C_{L}$: 100}\\
\hline
Sampling space&\multicolumn{4}{c|}{}&\multicolumn{4}{c|}{}  &\multicolumn{4}{c|}{}&\multicolumn{4}{c}{}\\
Gain (dB) / I (A) &\multicolumn{4}{c|}{[$40, 45$] / [$10^{-5}$, $10^{-4}$]}&\multicolumn{4}{c|}{[$10, 20$] / [$10^{-3}$, $10^{-2}$]}  &\multicolumn{4}{c|}{[$20, 30$] / [$10^{-4}$,$10^{-3}$]}&\multicolumn{4}{c}{[$40, 45$] / [$10^{-3}$,$10^{-2}$]}\\
% I (A)&\multicolumn{4}{c|}{[$10^{-5}$, $10^{-4}$]}&\multicolumn{4}{c|}{[$10^{-3}$, $10^{-2}$]}  &\multicolumn{4}{c|}{[$10^{-4}$,$10^{-3}$]}&\multicolumn{4}{c}{[$10^{-3}$,$10^{-2}$]}\\
PM ($\circ$) / BW (MHz)&\multicolumn{4}{c|}{[$>50$] / [$0.5$, $1$]}&\multicolumn{4}{c|}{[$>60$] / [$1$, $20$]} &\multicolumn{4}{c|}{[$>85$] / [$4$, $6$]} &\multicolumn{4}{c}{[$>55$] / [$1$, $2$]}\\
% BW (Hz) &\multicolumn{4}{c|}{[$5\times10^{5}$, $1\times10^{6}$]}&\multicolumn{4}{c|}{[$1\times10^{6}$, $2\times10^{7}$]} &\multicolumn{4}{c|}{[$4\times10^{6}$, $6\times10^{6}$]} &\multicolumn{4}{c}{[$1\times10^{6}$, $2\times10^{6}$]}\\
\bottomrule
\end{tabular}
\end{footnotesize}%scriptsize
\label{tab: tab1}
\vskip -15pt
\end{table*}

% \vskip -12pt

\subsection{Optimization Methods for Policy Training}
\label{sec: optimization_method}

% \iffalse
Combining the GAT and FCNN forms the policy network $\pi_{\theta}(a|s)$ parameterized by $\theta=\{W_{\text{GAT}}, W_{\text{FC}}\}$.
Here, $W_{\text{GAT}}$, $W_{\text{FC}}$ are the learnable parameters for the GAT and FCNN.
Our goal is to make the RL agent gain rich circuit design experiences and generate higher-quality decisions by interacting with the environment.
We can formally define the objective function of automated design of analog circuits as follows. 
\begin{equation}
J(\theta, G)={1}/{H} \cdot \sum\nolimits_{g\sim G}\mathbb{E}_{g,s\sim \pi_{\theta}}[R_{s,g}].
\label{eq:obj}
%    \vskip -6pt
\end{equation}
Here, $H$ is the the space size of all desired specifications $G$ and $R_{s,g}$ is the episode reward.
%defined in Eq.~\eqref{eq:reward_epi}. %Given each episode reward, 
Given the cumulative reward for each episode, we use Proximal Policy Optimization (PPO)~\cite{ppo} to update the parameters of the policy network with a clipped objective:
%shown below:
\begin{equation}
 L^{\text{CLIP}}(\theta) = \hat{\mathbb{E}}_k[\min(b_i(\theta), \text{clip}(b_k(\theta), 1-\epsilon, 1+ \epsilon ))\hat{A}_k],
\label{eq:ppo}
%    \vskip -6pt
\end{equation}
where $\hat{\mathbb{E}}_k$ represents the expected value at time step $k$; $b_k$ is the probability ratio of the new policy and the old policy, and $\hat{A}_k$ is the estimated advantage at time step $k$.

Previous RL-based methods~\cite{RL_1, RL2} for P2S tasks mainly explore Deep Deterministic Policy Gradients (DDPG) to train RL agents and have shown promising performance. 
However, the lack of a detailed comparison between different RL algorithms makes it difficult to determine which is better for P2S tasks.
DDPG is an off-policy RL method that uses two separate policies for exploration and updates, a stochastic behavior policy for exploration, and a deterministic policy for the target update.
The ``deterministic'' in DDPG refers to the fact that the agent computes the action directly instead of a probability distribution over actions.
DDPG is specifically designed for environments with continuous action spaces and continuous state spaces, making it an equally valid choice for continuous control tasks applicable to fields such as robotics or autonomous driving.

On the other hand, PPO is an on-policy RL method, that is, it involves collecting a small batch of experiences by interacting with the environment according to the latest version of its stochastic policy and using that batch to update its decision-making policy. 
The ``stochastic'' in PPO refers to the fact that the agent computes the action as a probability distribution instead of directly over actions.
PPO can often work with both discrete and continuous action spaces, making it suitable for a wide range of reinforcement learning tasks in various domains, e.g., training ChatGPT.
In particular, we use RL with discrete action space to build our framework due to: (i) experienced human designers also use fine-grained tuning (i.e., adjusting device parameters with several discrete tuning steps) to tackle the P2S task;
(ii) the thorough study in Section~\ref{sec: rl_algo} shows that PPO with discrete action space achieves better performance.

\section{Experimental Methodology}
\label{sec: exp_metho}

In this section, we present the experimental methodology for evaluating the proposed framework.
First, we introduce the circuit benchmarks used in our evaluations.
Then, baselines for comparisons are briefly discussed.
Finally, we show the training platform and configurations of our framework.

\subsection{Benchmarks and Performance Metrics for Evaluation}
\label{sec: circuit}

%\subsubsection{Circuit benchmarks}

Operational amplifiers (Op-Amps) are commonly used as circuit benchmarks in prior art~\cite{sv1, sv2, RL1, RL2, RL4, RA} and are also widely used as essential building blocks in many analog subsystems.
Therefore, we take multiple Op-Amps to evaluate the proposed framework.
In particular, we adopt a single-stage cascode Op-Amp, a two-stage Op-Amp, a folded-cascode Op-Amp~\cite{fco}, and a three-stage nested Miller compensation Op-Amp with feedforward transconductance stage~\cite{NMCF} (NMCF) in our benchmark.
These circuits have diverse topologies and design complexities.
The detailed schematics of these circuits were shown in previous work~\cite{sv1, sv2, RL1, RL2, RL4, RA} and are thereby omitted here.
The design space for the device parameters, the sampling space for the circuit specifications, and the PVT corners are listed in Table~\ref{tab: tab1}. 
There are $4\times2\times2 = 16$ extreme PVT conditions, including $4$ process variations, $2$ voltage variations, and $2$ temperature variations.

With the circuit benchmark, we examine mainly the sampling efficiency, design success rate, and design efficiency of our framework.
We show the sampling efficiency of RoSE-Opt by using a control experiment, that is, to train the RL agent with/without the BO vanguard.
The \textbf{sampling efficiency} is defined as the number of SPICE simulations saved to achieve the same training quality (i.e., training reward) compared to the control group.
To allow a reasonable comparison between different design automation methods, we also propose a $\text{FoM}_{\text{deploy}}$ defined as the ratio between the design success rate ($N_{\text{success}}$) and the design efficiency ($N_{\text{step}}\cdot T_{\text{sim}}$): $\text{FoM}_{\text{deploy}} = N_{\text{success}}/({N_{\text{step}}\cdot T_{\text{sim}}}$).
Here, $N_{\text{success}}$ is the \textbf{design success rate} of policy deployment by giving $200$ groups of design goals randomly sampled from the specification space. 
$N_{\text{step}}$ is the average number of required deployment steps (i.e., the number of circuit-level simulations) to achieve a group of design goals sampled from the specification space.
$T_{\text{sim}}$ is the simulation time for each simulation run at the circuit level. 
Note that the training time for learning-based methods is not included here, as it can be amortized during the deployment phase once the models are well trained (similar to the inference stage in supervised learning).

\subsection{Experiment Platform, Configurations, and Baselines}
Our framework is built on Python.
We create the circuit graph using the Deep Graph Library~\cite{dgl} and use Ray~\cite{ray}, a well-developed hyperparameter tuning package, to train RL agents. 
We implement all methods with PyTorch and BoTorch~\cite{botorch}.
All experiments were carried out on a 16-core Intel CPU.
We train separate RL agents for each circuit.
%\textcolor{blue}{For experiments that involve the BO vanguard to optimize the initial starting point, the maximum simulation budget for BO vanguard is 50 and the improvement threshold is set to be 0.002.}\footnote{\textcolor{blue}{Both hyperparameter values are set based on our empirical study. }}.
Note that we only need to run BO once at the beginning since we can reuse the starting point optimized by the BO vanguard in each RL's training and deployment phases. 
To achieve a more reliable and reproducible experiment result, we decided to run our BO vanguard 50 times and choose the starting point with a mean reward to minimize the variation caused by the initial random sampling. 
To provide detailed performance comparisons of different RL algorithms, we choose PPO~\cite{ppo} and DDPG~\cite{ddpg} as two representatives of the study and also use their default configurations to train policy networks.

Although various previous methods have been proposed to address P2S tasks, such as BO~\cite{BO}, Genetic Algorithm~\cite{genetic}, and RL methods~\cite{RL1, RL2}, they do not consider PVT variations and post-layout parasitics in the optimization process of device parameters.
\textcolor{black}{As a result, we modify all of these previous works to include all PVT corners for a direct comparison}. 
We also compare to the most recent work, RobustAnalog~\cite{RA}, which solves the P2S task considering the effect of partial variations in PVT.
Despite several major differences between RobustAnalog and RoSE-Opt, we care most about the efficacy of RobustAnalog in robust design, as it uses task pruning with reduced PVT corners for RL training, while our RL backbone considers all PVT corners.
We follow this strategy to implement RobustAnalog by modifying our RL backbone.

\section{Experimental Evaluations}
\label{sec:exp}

In this section, we show the evaluation results and compare the performance of our proposed framework with previous methods.
\textcolor{black}{First, we present the efficacy of incorporating domain knowledge into the learning loop}.
Second, we show our framework's sampling efficiency and robustness against PVT variations.
Third, we show our framework's capability to achieve reliable device sizing by taking into account post-layout parasitics.
Fourth, we show how the trained RL agent of our framework assists human designers in finding optimized device parameters, even if it fails in deployment in some cases.
Fifth, we present the performance of different RL algorithms in training RL agents for the P2S task.
\textcolor{black}{Finally, we also show the performance of our proposed method in Pareto optimization}.
The section ends with a summarization of comprehensive comparisons between our work and the prior arts.

\subsection{Efficacy, Robustness, and Sampling Efficiency}
\label{sec: robustness and sample efficiency}

\subsubsection{Efficacy of incorporating domain knowledge and customizing a multi-modal policy network}
We conducted comprehensive experimental studies to show the effectiveness of incorporating domain knowledge and customizing a multi-modal policy network. First, we present the importance of integrating domain knowledge into the learning process (e.g., taking PVT variations as an example). Existing methods~\cite{RL1,RL2} often overlook PVT variations and parasitic effects in the layout when designing reward functions, leading to failures during deployment in variable environments. In contrast, our proposed framework with a tailored variation-aware reward function achieves a design success rate that exceeds 90\% in non-ideal environments with PVT variations as listed in Table~\ref{tab: tab2}. Second, we show the superiority of customizing a multi-modal policy network. We compared different policy networks used by previous methods~\cite{RL1,RL2} with a consistent variation-aware reward function. Specifically, the work~\cite{RL1} uses a simple FCNN as a policy network and does not exploit circuit topologies. The work~\cite{RL2} introduces a GCN policy network based on a partial circuit topology and uses static semiconductor technology information as observations. Our study reveals that the GCN-based network could experience training divergence, necessitating adjustments based on our methodology (e.g., using node features with dynamic device parameters) for a reliable comparison. The comparisons between different methods are illustrated in Fig.~\ref{fig: fig_multimodal}. Our customized GNN-FCNN policy network demonstrates superior performance, yielding higher rewards with the two circuits.
In particular, the GAT-FCNN-based policy surpasses the GCN-FCNN-based policy, highlighting the effectiveness of our proposed policy network architecture. Unless otherwise specified, GAT-FCNN will be the default learning model for the remainder of this section, generally referred to as GNN-FCNN. These experimental studies demonstrate the effectiveness of integrating domain knowledge and tailoring a multi-modal policy network.

\begin{figure}[!t]
 % \vskip -0.1in
\begin{center}
\includegraphics[width=1.0 \linewidth]{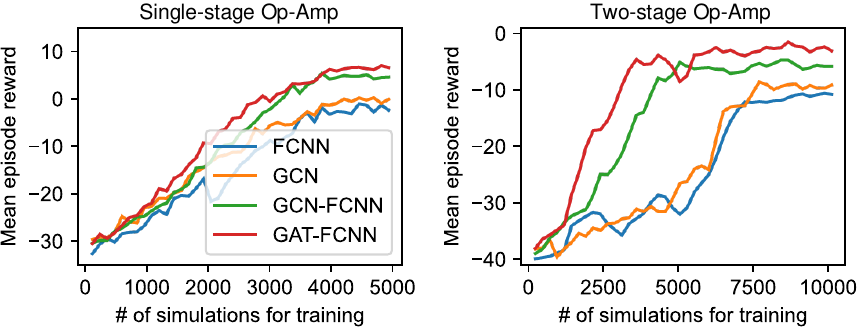}
\vskip -3pt
\caption{A comparison between our customized policy network and prior methods' policy networks (GCN~\cite{RL2}, FCNN~\cite{RL1}) for the variation-aware P2S task.}
\label{fig: fig_multimodal}
\end{center}
 \vskip -15pt
\end{figure}

\subsubsection{Robust design with full PVT incorporation} We then show the robust design enabled by our RL backbone against PVT variations by deploying our RL agent in an environment taking full account of PVT variations.
Policy deployment applies a trained policy to automatically find the device parameters for given design goals.
The left column of Fig.~\ref{fig: fig_4} shows the deployment trajectories under several representative PVT corners by taking the phase margin of the Folded-cascode Op-Amp as an example, where each color represents a PVT corner.
It can be seen that although each trajectory under a specific PVT corner is smooth, the worst corner can be quickly replaced by another corner due to the competition between different corners.
Here, the worst case indicates the corner where the circuit specification deviates the most from the design goal.
This phenomenon shows that device sizing with PVT variations is much more complex compared to the nominal case.
Notably, by incorporating PVT variations into our method, our RL agent can achieve a robust design by finding optimal device parameters that can satisfy the design goal in all PVT corners.

To investigate the impact of this corner competition on actual RL training, we train three RL backbones with different levels of PVT incorporation. As shown in the right column of Fig.~\ref{fig: fig_4}, although the RL backbone trained with the nominal corner (or partial PVT corner) achieves convergence with high reward, it completely (or partially) diverges when deployed in an environment with full PVT variations (i.e., the dashed curves). The result verifies that device sizing with full PVT variations is much more complex compared to the nominal case. It also suggests that previous work, RobustAnalog~\cite{RA}, which uses the K-means clustering to prune the PVT corners during the training process, is not sufficient to ensure robustness in an environment with full variations of PVT. It can delay the training process because the RL algorithm needs more training steps to repeat the clustering whenever the device parameters reach the design goal under partial PVT corners, but fail in the setup of full PVT corners.

\begin{figure}[!t]
% \vskip -0.20in
\begin{center}
\centerline{\includegraphics[width=1.0\linewidth]{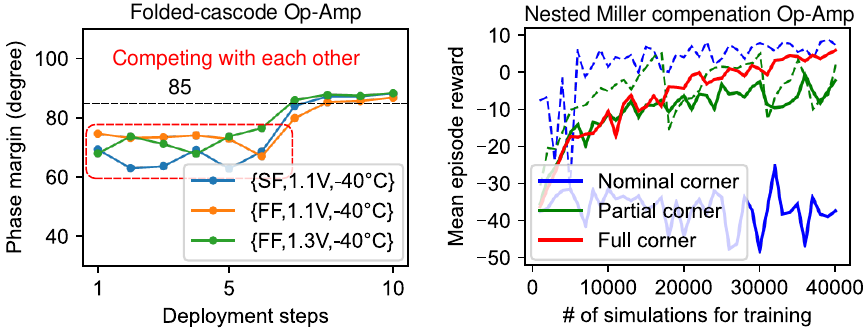}}
\vskip -3pt
\caption{Left: illustration of the competing phenomenon between PVT corners. Right: comparison between different levels of PVT incorporation for RL backbone training. Dashed lines are the actual nominal/partial corner training curve. Solid lines are the training curve evaluated under full PVT corners.}
\label{fig: fig_4}
\end{center}
 \vskip -15pt
\end{figure}

\begin{figure}[!t]
 % \vskip -0.1in
\begin{center}
\includegraphics[width=1.0 \linewidth]{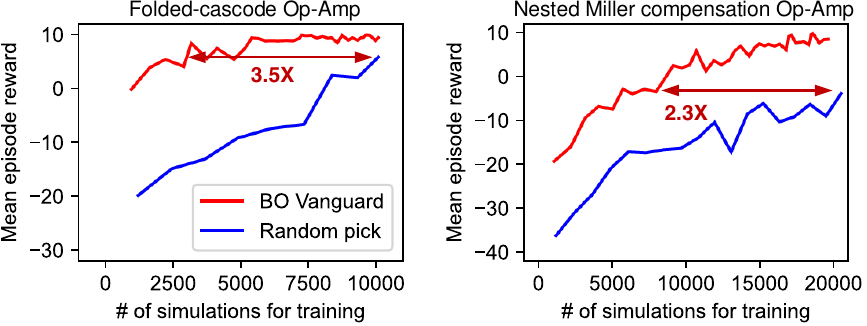}
\vskip -3pt
% \caption{An example to show the comparison of the training sample efficiency by using the RL Backbone with or without pre-optimization of BO.}
\caption{\textcolor{black}{An example to show the comparison of the training sample efficiency by using the RL Backbone with or without pre-optimization of its starting state.}}
\label{fig: fig_3}
\end{center}
 \vskip -15pt
\end{figure}

\subsubsection{Efficient sampling with BO Vanguard}
Thirdly, we show that the BO vanguard can improve sampling efficiency to train our RL backbone.
Fig.~\ref{fig: fig_3} illustrates the example training curves of our RL backbone to design two types of Op-Amps with two different starting points, one is from BO searching (labeled as ``BO Vanguard'') and the other is a randomly selected value from the device parameter design space $S_{P}$, e.g., median value (labeled as ``Random pick'').
It shows that the RL agent without an optimized starting point often needs more circuit-level simulations to achieve the same reward as the one with an optimized starting point from BO (e.g., in this case, $3.5\times$ for Folded-cascode Op-Amp and $2.3\times$ for three-stage nested Miller compensation Op-Amp).
Thus, by optimizing the starting point, the RL agent converges faster with fewer sampling data (that is, fewer circuit-level simulations). 

\begin{table*}[!hbt]

\caption{Comparison of device parameters before and after physical design with design goals of gain = 45 dB, bandwidth = 800 kHz, phase margin = 50$^{\circ}$, and power = 15 $\mu$W.}
% \vskip -6pt
\begin{center}
\begin{footnotesize}
\vskip -0.15in
\begin{tabular}{c|c|c|c|c|c|c|c|c|c|c|c}
\hline
Device           & mp1     & mp2     & mp3     & mp4     & mn1     & mn2     & mn3     & Gain      & BW      & PM              & Power  \\ \hline
Schematic design & 1.02 $\mu$m & 1.02 $\mu$m & 1.02 $\mu$m & 1.02 $\mu$m & 6.2 $\mu$m & 4.94 $\mu$m & 6.2 $\mu$m & 47.3 dB & 862 kHz & 52$^{\circ}$ & 12 $\mu$W \\ \hline
Layout design    & 1.02 $\mu$m & 1.02 $\mu$m & 1.02 $\mu$m & 1.02 $\mu$m & 5.26 $\mu$m & 5.26 $\mu$m & 5.26 $\mu$m & 46.91 dB & 897 kHz & 53.52$^{\circ}$ & 12 $\mu$W \\ \hline
\end{tabular}
\end{footnotesize}
\end{center}
\label{tab: tab_post}
\vskip -12pt
\end{table*}

\subsection{Parasitic-Aware Device Parameter Optimization}
%\subsubsection{Parasitic-Aware Device Parameter Optimization(change order)}
\label{sec: para_sizing}

We continue to study how to apply RoSE-Opt to optimize parasitic-aware device parameters.
Without considering the parasitic effect of physical layouts at the pre-layout design stage, the obtained device parameters cannot guarantee the circuit specifications after the schematic is directly transferred into a physical design.
In practice, there are often tens of iterations between schematic design and physical design performed by human designers to fine-tune the device parameters to ensure that the circuit under design meets the design goals. 

\textcolor{black}{Several previous works have explored learning-based methods to address this parasitic-aware optimization problem~\cite{RL1,para_aware,budak2023practical}}.
An early RL-based method~\cite{RL1} aims to tackle it by deploying the trained RL agent in a parasitic-aware environment.
In particular, this method uses the BAG tool~\cite{BAG} to automatically generate a physical layout for the circuit based on the device parameters at each deployment step and exploits the reward from post-layout simulation to guide the search process for the trained RL agent.
This process continues until the agent meets the target when parasitics are considered or it has reached the maximumly allocated deployment steps. 
Another work~\cite{para_aware} attempts to tackle the problem by combining supervised learning and BO.
The idea is to train a graph neural network to predict the parasitics of an analog circuit with its device parameters and then back-annotate the parasitics to the circuit schematic.
With this processing step, BO is applied to search for optimal device parameters through the parasitic-aware schematic.

These previous efforts have shown good performance in finding reliable device parameters to meet design goals after post-layout simulation.
However, the physical design of analog circuits is quite flexible.
Even for the same circuit, different human designers can construct different physical layouts.
The BAG tool is limited to generating a few fixed layouts for some typical circuits.
Training a GNN to predict parasitics requires a huge amount of data and suffers from approximation error.
Therefore, the previous methods do not apply to general cases.

\begin{figure}[!t]
%\vskip -0.05in
\begin{center}
\includegraphics[width=0.8\linewidth]{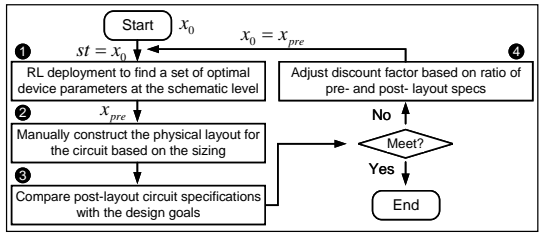}
\vskip -3pt
\caption{The flow RoSE-Opt used for the parasitic-aware sizing. Please refer to Fig.~\ref{fig: human_loop} for the parasitic aware experiment's schematic and layout.
}
\label{fig: para}
\end{center}
\vskip -18pt
\end{figure}

We explore another method to solve the same problem with much higher flexibility.
Our method relies on two key observations from the human design loop.
First, human experts often construct an initial physical layout of the circuit with an initial set of device parameters and fine-tune the device parameters by following the same placement of the device as the one used in the initial physical layout. 
Second, circuit specifications from the post-layout simulation of this initial physical layout are often degraded compared to the desired goals but are not far from them.
Therefore, the optimal final device parameters to meet the design goals also fall in the neighborhood of the initial set of device parameters.

With these key observations in mind, our method can apply to parasitic-aware device parameter optimization by following the essential steps shown in Fig.~\ref{fig: para}.
We begin by initializing all discount factors to 1. 
These discount factors are explained in \circlednew{white}{4}.
The other steps are as follows:

\noindent\circlednew{white}{1} Deploy the trained RL agent to find the set of optimal device parameters that satisfy the design goals in the pre-layout stage and perform simulations to obtain the circuit specifications with this set of device parameters; the $i^{\text{th}}$ specification of the $j^{\text{th}}$ corner is marked as $s_{i,\text{pre}}^j$.

\noindent\circlednew{white}{2} Construct a physical layout with the device parameters found in Step \circlednew{white}{1}.

\noindent\circlednew{white}{3} Extract the circuit specifications (i.e., $s_{i,\text{post}}^j$) of this physical layout by performing a post-layout simulation and compare them with the design goals; if satisfied, the design is successful; otherwise, jump to Step \circlednew{white}{4}.

\noindent\circlednew{white}{4} Adjust the discount factor based on the ratio between pre- and post- layout specifications, e.g., if $s_{i,\text{pre}}^j \geq s_{i,\text{post}}^j$, $\alpha_i^j=s_{i,\text{post}}^j/s_{i,\text{pre}}^j$. 

After the first iteration, we repeat the flow using the set of device parameters $x_{pre}$ found in Step \circlednew{white}{1} as the new starting point $st$ of the trained RL agent for another deployment and the intermediate circuit specification in \circlednew{white}{1} will be discounted by the discount factor as $\alpha_i^j \cdot s_{i,\text{pre}}^j$, until the optimal final device parameters that satisfy the design goals in the pre-layout stage are found. This method essentially follows a dynamic over-design strategy that can alleviate the challenge of estimating a static over-design value derived from domain knowledge for complex circuits. Our experiments with the single-stage Op-Amp illustrated in Fig.~\ref{fig: human_loop} show that our method generally takes no more than two rounds to reach a set of device parameters with which the circuit specifications of a physical layout can also meet the design goal. Table~\ref{tab: tab_post} shows optimized device parameters with/without consideration of parasitic effects. The final physical layout is similar to that shown in Fig.~\ref{fig: human_loop} and is omitted here.

\subsection{Analysis of Failed Deployment Cases}
%\subsubsection{Analysis of failed deployment cases}

\label{sec: deployment}
\begin{figure}[!t]
%\vskip -0.05in
\begin{center}
\includegraphics[width=0.8\linewidth]{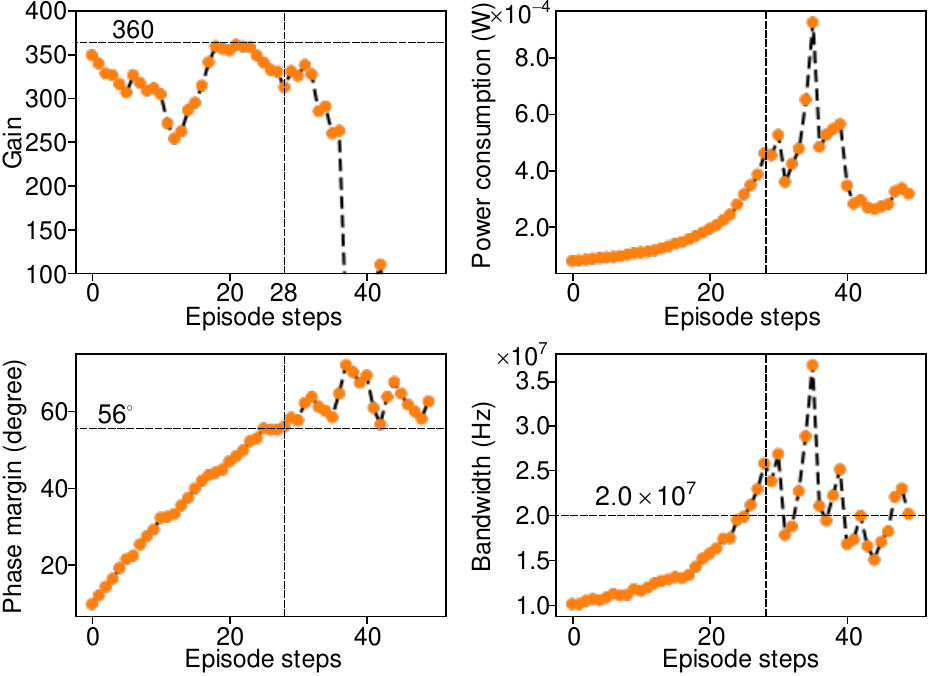}
\vskip -3pt
\caption{Failed policy deployment in the two-stage Op-Amp example. The highest reward appears in the $28^{\text{th}}$ step. After slight manual adjustment from that step, a set of optimal device parameters can often be easily obtained as shown in Table~\ref{tab_man}.
}
\label{fig: fail}
\end{center}
\vskip -12pt
\end{figure}

\begin{table}[!t]
% \centering
\caption{Detailed device parameters during the policy deployment for the two-stage Op-Amp.}
\begin{center}
\begin{footnotesize}
\vskip -0.15in
\begin{tabular}{c|c|c|c|c}
\toprule
% \hline
\textbf{Parameter} & \textbf{Step 27} & \textbf{Step 28} & \textbf{Step 29} & \textbf{Manual tuning} \\
\hline
mpl ($\mu$m) & 11 & 10 & 10 & 10 \\
\hline
mn1 ($\mu$m) & 35 & 34 & 34 & 35 \\
\hline
mp3 ($\mu$m) & 81 & 83 & 85 & 83 \\
\hline
mn3 ($\mu$m) & 17 & 16 & 15 & 17 \\
\hline
mn4 ($\mu$m) & 5 & 4 & 4 & 4 \\
\hline
mn5 ($\mu$m) & 45 & 47 & 49 & 47 \\
\hline
c1 (pF) & 4.4 & 4.3 & 4.5 & 3.6 \\
\hline
Reward & -0.105 & -0.057 & -0.071 & 10 \\
\bottomrule
\end{tabular}
\end{footnotesize}
\end{center}
\label{tab_man}
\vskip -0.25in
\end{table}

Our trained RL agent achieves a high design success rate with policy deployment (i.e., $>$90\% across different circuits as reported by our prior work~\cite{rose}).
We find that for these failed cases, some circuit specifications are able to reach the design goals, while the others converge to a neighborhood of the desired ones at some deployment steps, but after which they deviate a bit from the goals.
Fig.~\ref{fig: fail} shows such a failed policy deployment in the two-stage Op-Amp example, where the desired circuit specifications given are gain ($G=360$), bandwidth ($B=2.0\cdot 10^7$ Hz), phase margin ($PM= 56^{\circ}$), power consumption ($P=6.93\cdot 10^{-3}$ W).
It is observed that around the $28^{\text{th}}$ step, the bandwidth, phase margin, and power consumption are satisfied, but the gain is still lower than the design goal. 
We examine the detailed device parameters\footnote{Note that here we do not show the finger numbers of transistors because they generally remain unchanged around the $28^{\text{th}}$ step.} around the $28^{\text{th}}$ step as shown in Table~\ref{tab_man}.
It shows that the reward achieves the highest value in the $28^{\text{th}}$ step.
We then select the device parameter values reached at this step and proceed with a slight manual tuning starting from these values.
With less than five manual tuning iterations, a set of optimal device design parameters can often be easily obtained\footnote{Note that due to the limitations of current deep learning techniques, learning-based methods (along with other approaches) cannot guarantee a 100\% success rate. Addressing these failures requires the intervention of human designers.
We anticipate that future advances in deep learning will address this issue.}.
The last row of Table~\ref{tab_man} shows the device parameters obtained after slight manual adjustment.

\subsection{Comparisons between Different RL Algorithms}
%\subsubsection{Comparisons between Different RL Algorithms}
\label{sec: rl_algo}

Different RL algorithms have shown different performance in solving practical problems.
PPO and DDPG are two primary RL algorithms used in current RL-based methods~\cite{RL1, RL_1, RL2} for P2S tasks.
Here, we perform detailed experiments to compare their performance in tackling the P2S task by using the design of a two-stage Op-Amp as an example.
Fig.~\ref{fig: rl_algorithm comparison} illustrates our evaluation results, where we train RL agents with PPO using both ``discrete'' and ``continuous'' actions, as well as DDPG with ``continuous'' action.
Each curve in Fig.~\ref{fig: rl_algorithm comparison} is based on 6 random seeds with $3\times10^{6}$ simulations\footnote{While training results for other methods are based on millions of simulations, our RoSE-Opt method achieves rapid convergence and high design success rates within just tens of thousands of simulations, indicating its practical promise.} budget for training. \textcolor{black}{All the RL algorithms perform batch training by collecting 6 simulation results in parallel.} Note that we also observe similar results for other types of Op-Amps.

\subsubsection{PPO-continuous vs. DDPG-continuous}
Compared to DDPG-continuous, we find that PPO-continuous has a lower sample efficiency during the training process.
This is because PPO-continuous adopts an on-policy learning mechanism that samples actions according to its latest stochastic policy.
The on-policy characteristic introduces variance, since each estimate of an expectation over a finite set of samples may vary, which necessitates a large number of samples for accurate mean calculations, thereby leading to low sampling efficiency.
In contrast, DDPG-continuous utilizes an off-policy learning mechanism, which involves a replay buffer to store transitions from the previous policy and relies on the current policy only to replenish the buffer, improving the sampling efficiency.
The lower sampling efficiency of PPO-continuous also impacts the training quality of the policy.
As shown in Fig.~\ref{fig: rl_algorithm comparison}, with the same number of training samples, the mean design simulation cost (design accuracy) of the trained policy with PPO-continuous (green line) is higher (lower) than that of the one with DDPG-continuous (red line).

\begin{figure}[!t]
% \vskip -0.1in
\begin{center}
\includegraphics[width=1.0\linewidth]{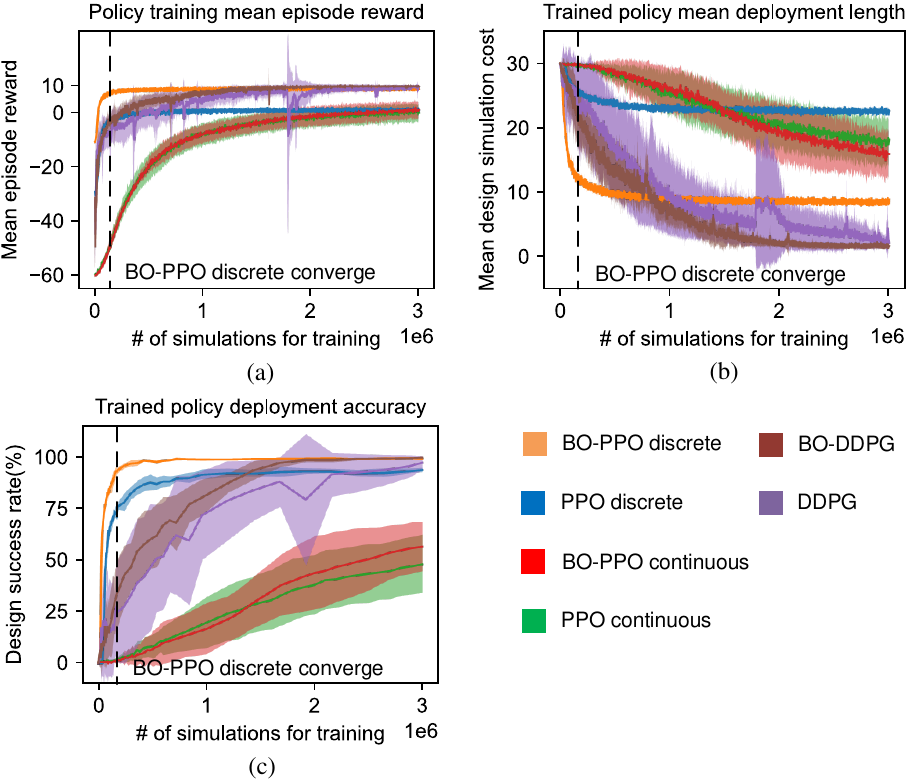}
\vskip -3pt
\caption{Comparisons between DDPG and PPO when applied to train RL agents and deploy trained checkpoints for 200 randomly sampled design goals. (a) mean episode reward for training, (b) mean deployment simulation cost (c) mean design success rate. Each curve is based on 6 random seeds. The vertical dashed line is the convergent point for BO-PPO discrete (our method).}
\label{fig: rl_algorithm comparison}
\end{center}
\vskip -21pt
\end{figure}

\subsubsection{PPO-discrete vs. DDPG-continuous}

With the setting of discrete action space, PPO-discrete demonstrates superior training sample efficiency and more consistent results compared to its continuous counterpart and DDPG-continuous.
In a discrete environment, the action choices of an RL agent at each step are simplified to adjust the parameter upward/downward with a small increment/decrement, or to maintain its current value. 
This simplicity makes the training process smoother and improves the quality of the policy to find the optimal device parameters during the deployment process, as shown in Fig.~\ref{fig: rl_algorithm comparison} (blue line).
In a complex continuous environment where the agent has a plethora of parameter choices, the off-policy mechanism of DDPG-continuous could suffer from biases, where some of the updates are based on prior (potentially incorrect) expectation estimates. 
This leads to irreversible incorrect estimates in the end and causes training instability due to its inherent low-variance but high-bias nature.
As shown in Fig.~\ref{fig: rl_algorithm comparison} (red line), there is a sudden change with respect to the mean episode reward and design accuracy/cost.

However, there is a caveat to using PPO-discrete: the discrete environment constrains the policy's design efficiency. 
Its mean design cost (i.e., simulation steps) is influenced by the granularity of the step in the discrete space and the distance between the initial state and the target solution.
Thus, more design steps are required to find the optimal device parameters.
In a continuous environment, the action at each step corresponds to the normalized device parameters in the design space, thereby demanding fewer design steps.

\subsubsection{BO-PPO-discrete vs. PPO-discrete}
To address the issue of the low design efficiency of PPO-discrete, we introduce BO to optimize the starting point of PPO-discrete, thus positioning the initial state closer to the solution space. 
As illustrated in Fig.~\ref{fig: rl_algorithm comparison} (yellow line), this strategy could optimize the starting point, allowing PPO-discrete to reach the desired specifications without too much meaningless exploration, accelerating the trajectory formulation towards the solution.
Ultimately, this integration of BO with PPO-discrete, termed BO-PPO discrete, demonstrates superior performance in both design accuracy and sampling efficiency, and achieves commendable results in design efficiency. For other continuous BO-RL algorithms (i.e., BO-DDPG-continuous vs. DDPG-continuous), we also observe performance improvement compared to its RL baseline by producing more consistent results. This novel approach showcases the potential of combining classical optimization techniques to improve the effectiveness of RL in complex analog circuit design. As optimization techniques continue to evolve, the combined strengths of different optimization methods, such as BO and RL, underscore the importance of hybrid strategies in complex optimization scenarios of analog circuit parameters.

\subsubsection{Training Reward Is Not Always A Good Metric for Comparing Different RL Policies}
%\noindent{\textbf{Training Mean Episode Reward Is Not Always A Good Metric for Comparing Different RL Policies:}} 
One last thing to note is that DDPG-continuous is able to achieve a much larger episode reward compared to PPO-discrete during the training process but suffers from worse design accuracy during the deployment process. 
This discrepancy is due to the nature of the episode reward function, which accumulates intermediate rewards throughout the search process for each training episode.
PPO-discrete leverages fine-grained action to search for optimal device parameters through multiple steps.
Due to this fine-grained mechanism, the improvement of intermediate rewards in a training episode is slow, resulting in a smaller episode return.
In contrast, DDPG-continuous adopts continuous action, allowing it to find a suboptimal solution earlier in the search process and even within a single step, thereby leading to a larger accumulated reward.
However, since it suffers from biases as discussed above, DDPG-continuous does not achieve a high design success rate in the deployment stage.
This finding shows that leveraging the training reward as a metric to compare design automation methods based on different RL algorithms, as done in many previous work~\cite{RL_1, RL2}, could be misleading.
We recommend employing deployment accuracy and design efficiency as metrics for fair and reasonable comparisons across various methods.

\begin{table*}[ht]
  \centering
  %\vskip -0.6in
\captionof{table}{Summary of comparison with existing learning and optimization methods.}
\vskip -6pt
\begin{footnotesize}%scriptsize
\begin{threeparttable}
\begin{tabular}{c|cc|c|cccc|cc}
\toprule
\multicolumn{1}{c|}{}&\multicolumn{2}{c|}{Domain}&\multicolumn{1}{c|}{Starting}&\multicolumn{4}{c|}{P2S problem}&\multicolumn{2}{c}{$\text{FoM}_{\text{Op-Amp}}$ opt problem} \\
\cline{5-10}
\multicolumn{1}{c|}{Methods}&\multicolumn{2}{c|}{knowledge}&\multicolumn{1}{c|}{point}&\multicolumn{2}{c|}{Training sample efficiency} & \multicolumn{2}{c|}{$\text{FoM}_{\text{deploy}}$}& \multicolumn{2}{c}{$\text{FoM}_{\text{Op-Amp}}$ value} \\
\cline{2-3}
\cline{5-10}
&\multicolumn{1}{c|}{Multimodal}&\multicolumn{1}{c|}{PVT}&\multicolumn{1}{c|}{optimization}&  \multicolumn{1}{c|}{folded-cascode} & \multicolumn{1}{c|}{NMCF}& \multicolumn{1}{c|}{folded-cascode} & \multicolumn{1}{c|}{NMCF}& \multicolumn{2}{c}{NMCF}\\
\hline
%\multicolumn{1}{c|}{RobustAnalog}& \multicolumn{1}{c|}{No} & \multicolumn{1}{c|}{Yes} & \multicolumn{1}{c|}{N/A} & \multicolumn{1}{c|}{N/A} & \multicolumn{1}{c|}{N/A} & \multicolumn{1}{c}{N/A}\\
%\hline
\multicolumn{1}{c|}{RobustAnalog\tnote{a} ~\cite{RA}} & \multicolumn{1}{c|}{Yes} & \multicolumn{1}{c|}{Partial}
&\multicolumn{1}{c|}{No}
& \multicolumn{1}{c|}{1$\times$} & \multicolumn{1}{c|}{1$\times$} & \multicolumn{1}{c|}{1.14} & \multicolumn{1}{c|}{0.625}& \multicolumn{2}{c}{0.814}\\
\hline

\multicolumn{1}{c|}{BO\tnote{b} ~\cite{BO}} & \multicolumn{1}{c|}{No} & \multicolumn{1}{c|}{Full}
&\multicolumn{1}{c|}{N/A\tnote{c}}
& \multicolumn{1}{c|}{N/A\tnote{c}} & \multicolumn{1}{c|}{N/A\tnote{c}} & \multicolumn{1}{c|}{0.46} & \multicolumn{1}{c|}{0.044}& \multicolumn{2}{c}{2.935}\\
\hline

\multicolumn{1}{c|}{GA\tnote{b} ~\cite{genetic}} & \multicolumn{1}{c|}{No} & \multicolumn{1}{c|}{Full} 
&\multicolumn{1}{c|}{N/A\tnote{c}}
& \multicolumn{1}{c|}{N/A\tnote{c}} & \multicolumn{1}{c|}{N/A\tnote{c}} & \multicolumn{1}{c|}{0.067} & \multicolumn{1}{c|}{0.005}& \multicolumn{2}{c}{1.054}\\
\hline

\multicolumn{1}{c|}{RL1\tnote{b} ~\cite{RL1}} & \multicolumn{1}{c|}{No} & \multicolumn{1}{c|}{Full}
&\multicolumn{1}{c|}{No}
& \multicolumn{1}{c|}{1.3$\times$} & \multicolumn{1}{c|}{1.32$\times$} & \multicolumn{1}{c|}{1.24} & \multicolumn{1}{c|}{0.9}& \multicolumn{2}{c}{1.378}\\
\hline

\multicolumn{1}{c|}{RL2\tnote{b} ~\cite{RL2}} & \multicolumn{1}{c|}{No} & \multicolumn{1}{c|}{Full}
&\multicolumn{1}{c|}{No}
& \multicolumn{1}{c|}{1.35$\times$} & \multicolumn{1}{c|}{1.4$\times$} & \multicolumn{1}{c|}{1.25} & \multicolumn{1}{c|}{1.1}& \multicolumn{2}{c}{1.436}\\
\hline

\multicolumn{1}{c|}{RL Backbone} & \multicolumn{1}{c|}{Yes} & \multicolumn{1}{c|}{Full}
&\multicolumn{1}{c|}{No}
& \multicolumn{1}{c|}{2$\times$} & \multicolumn{1}{c|}{2.5$\times$} & \multicolumn{1}{c|}{1.42} & \multicolumn{1}{c|}{1.26}& \multicolumn{2}{c}{1.664}\\
\hline

\multicolumn{1}{c|}{\textbf{RoSE-Opt}} & \multicolumn{1}{c|}{Yes} & \multicolumn{1}{c|}{Full} 
&\multicolumn{1}{c|}{\textbf{Yes}}
& \multicolumn{1}{c|}{\textbf{12$\times$}} & \multicolumn{1}{c|}{\textbf{7.9$\times$}} & \multicolumn{1}{c|}{\textbf{20.51}} & \multicolumn{1}{c|}{\textbf{6.87}}& \multicolumn{2}{c}{\textbf{3.812}}\\
% \hline
% \multicolumn{1}{c|}{Bayesian Optimization\tnote{b} ~\cite{BO}}& \multicolumn{1}{c|}{No} &\multicolumn{1}{c|}{N/A\tnote{c}}
% & \multicolumn{1}{c|}{N/A\tnote{c}} & \multicolumn{1}{c|}{N/A\tnote{c}} & \multicolumn{1}{c|}{0} & \multicolumn{1}{c}{0}\\
% \hline
% \multicolumn{1}{c|}{RL baseline\tnote{b} ~\cite{RL_1}}& \multicolumn{1}{c|}{No} &\multicolumn{1}{c|}{No}
% & \multicolumn{1}{c|}{Fail} & \multicolumn{1}{c|}{Fail} & \multicolumn{1}{c|}{0} & \multicolumn{1}{c}{0}\\
\bottomrule
\end{tabular}
\begin{tablenotes}\scriptsize
\item[a] Original work~\cite{RA} only uses normal 4-layer multilayer perceptions. We modify its network to the same multi-modality network we used in our RL backbone.
\item[b] Previous methods~\cite{BO,genetic,RL1,RL2} without considering PVT variations fail in the robust design with zero design accuracy and close to 0 $\text{FoM}_{\text{Op-Amp}}$. As a result, we modify their work to include full PVT incorporation.
\item[c] Traditional optimization methods~\cite{BO,genetic} do not need pre-optimized starting points and training (BO~\cite{BO} is still considered as an ad-hoc optimization method since it only uses active learning for improving its optimization sample efficiency without generalization to unseen design goals). Thus, some metrics are not applicable here. 
\end{tablenotes}
\end{threeparttable}
\end{footnotesize}
\label{tab: tab2}
\vskip -12pt
\end{table*}

\begin{figure}[!t]
%\vskip -0.05in
\begin{center}
\includegraphics[width=1.0\linewidth]{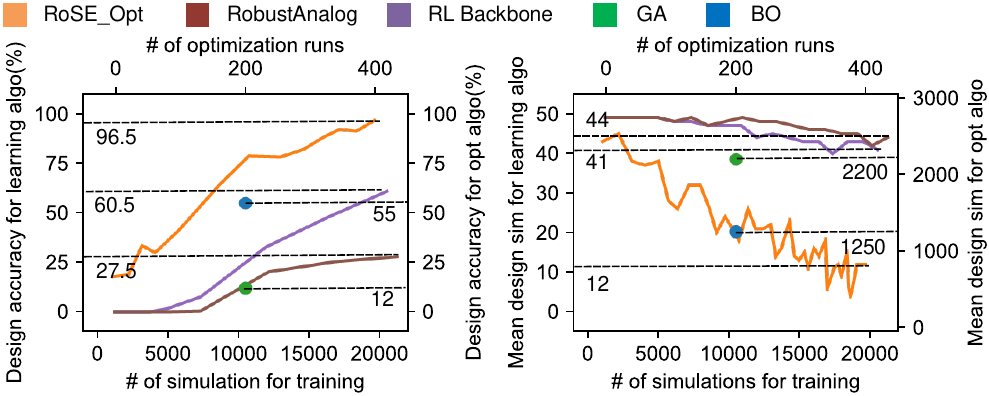}
\vskip -3pt
\caption{Comparisons between our proposed framework RoSE-Opt, our RL Backbone, RobustAnalog~\cite{RA}, BO~\cite{BO}, and GA~\cite{genetic} for variation-aware device sizing problem by taking the design of nested Miller compensation Op-Amp as an example. Left: mean design success rate. Right: mean design simulation cost.}
\label{fig: com}
\end{center}
\vskip -15pt
\end{figure}

\begin{figure}[!t]
%\vskip -0.05in
\begin{center}
\includegraphics[width=1.0\linewidth]{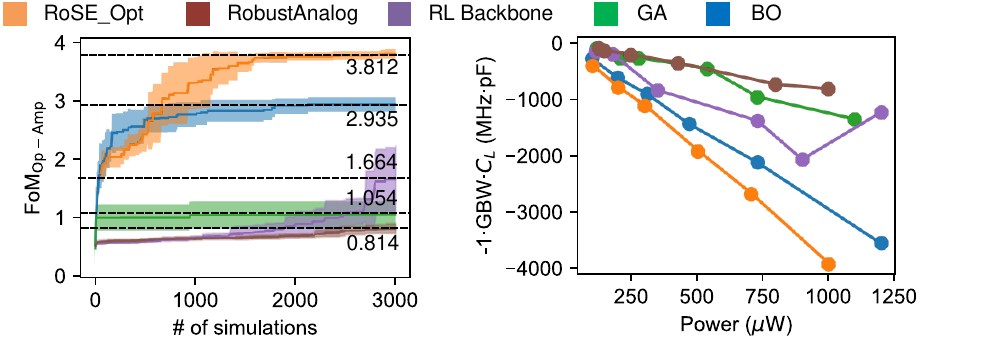}
\vskip -3pt
\caption{Comparisons between our proposed framework RoSE-Opt, our RL Backbone, RobustAnalog~\cite{RA}, BO~\cite{BO}, and GA~\cite{genetic} for Pareto optimization problem by taking the design of nested Miller compensation Op-Amp as an example. Left: $\text{FoM}_{\text{Op-Amp}}$ result. Right: Pareto frontier. Each curve is based on 6 random seeds.}
\label{fig: com_FoM}
\end{center}
\vskip -15pt
\end{figure}

\subsection{Comparisons between Different Methods and Summary}
\label{sec: sum_pre}
Finally, we compare our proposed RoSE-Opt method with previous RL methods~\cite{RL1,RL2,RA} and conventional optimization methods, such as BO~\cite{BO} and Genetic Algorithm~\cite{genetic}. In addition to a general variation-aware device sizing task that we introduced in Section~\ref{sec:RL}, we also study a specific task of analog design, i.e., Pareto optimization. The goal is to optimize a particular circuit specification or a figure-of-merit (FoM) that consists of multiple circuit specifications~\cite{cao2022domain}. Here, we use the standard FoM of Op-Amp defined as $\text{FoM}_{\text{Op-Amp}} \text{(MHz} \cdot \text{pF/ } \mu \mathrm{W}) = (\mathrm{GBW} \cdot C_L) / P$ to customize a reward function and train RL agents (or use BO/Genetic Algorithm) to maximize it. \textcolor{black}{Note that we also observe similar results for the nominal corner experiment.}

\subsubsection{Variation-aware device sizing}
Fig.~\ref{fig: com} illustrates the results.
With the benchmark circuit, RoSE-Opt can achieve a design success rate of 96.5\% and an average deployment length of 12 simulation costs within 20000 simulations\footnote{Our method aims to achieve high generality with learning-based methods and thus demands more data for training compared to conventional ad-hoc optimization methods such as BO.} for training\footnote{For P2S problem, the total optimization time is dominated by simulation time. The average simulation time of nested Miller compensation Op-Amp is around 10 seconds for each run. RoSE-Opt takes 20000 simulations to train its RL agent. After the training, RoSE-Opt takes an average of 12 simulation costs (2 minutes) for one unseen design goal.}.
In contrast, with the same number of simulations, the RL Backbone attains a lower design success rate and a longer deployment length.
The comparison again shows that the BO vanguard indeed improves sample efficiency.
RobustAnalog is even less competitive to the RL Backbone due to the incorporation of only partial PVT variations into the learning framework. Traditional ad-hoc optimization methods such as BO and GA cannot solve the variation-aware device sizing problem with high accuracy and efficiency. Thus, our RoSE-Opt framework achieves the best sampling efficiency, design efficiency, and design success rate by considering all PVT variations in the learning loop and adopting the two-level optimization method at the same time.

% \textcolor{red}{show a training time discussion here.}

\subsubsection{Pareto optimization}
Fig.~\ref{fig: com_FoM} shows the Pareto frontier of different methods.
Compared to baselines, our RoSE-Opt achieves the highest $\text{FoM}_{\text{Op-Amp}}$ and defines the optimal Pareto frontier by achieving the lowest power consumption while maintaining the highest GBW within the fewest simulations.

In the end, we summarize RoSE-Opt together with previous comparisons in Table~\ref{tab: tab2}. 
In particular, we use BO~\cite{BO} and GA~\cite{genetic} as representatives of prior optimization-based methods and use RobustAnalog~\cite{RA}, RL1~\cite{RL1}, and RL2~\cite{RL2} as representatives of previous learning-based methods. 
Additionally, we use the $\text{FoM}_{\text{deploy}}$ defined previously to evaluate the overall performance of a design automation method in the P2S problem and use the $\text{FoM}_{\text{Op-Amp}}$ for Pareto optimization evaluation. 
The comparisons show that our proposed RoSE-Opt framework, which uses BO as pre-optimization and a domain-knowledge-infused RL agent as the main optimization backbone, can significantly improve training sample efficiency, design efficiency, and design success rate for the PVT-aware design and improve Pareto optimization. 
Note that BO is performed only once at the very beginning, thereby incurring minimal overhead. In summary, our knowledge-infused RoSE-Opt framework that takes advantage of the complementary benefits of learning and optimization algorithms (i.e., combining BO and RL) can achieve the best FoM for the challenging reliable device sizing problem.

% with BO as pre-optimization and a domain knowledge-infused RL agent as the main learning or optimization backbone, 

\iffalse
The comparisons show that with BO as pre-optimization and a domain knowledge-infused RL agent as the main learning or optimization backbone, our RoSE-Opt framework can significantly improve training sample efficiency, design efficiency, and design accuracy for the PVT-aware design and improve Pareto optimization. 
Note that BO is performed only once at the very beginning, thereby incurring minimal overhead. In summary, our knowledge-infused RoSE-Opt framework that takes advantage of the complementary benefits of learning and optimization algorithms (i.e., combining BO and RL) can achieve the best FoM for the challenging reliable device sizing problem.

\fi

\section{Conclusion}
\label{sec:conc}
% We propose a BO-RL-based framework to automate the P2S task for the design of analog circuits.
% \textcolor{blue}{The key properties of our framework are: 1) incorporating domain knowledge of practical analog circuit design (e.g., the underlying physical topology of a given circuit, the trade-offs between specifications, PVT variations, and parasitic effects of physical layout) into the learning loop for ensuring a robust and optimal design. 2) leveraging the rapid convergence of BO to establish an optimized starting point for RL, thereby enhancing RL's training and deployment efficiency.}
\textcolor{black}{We propose a BO-RL-based framework to automate the P2S task for analog circuit design, by incorporating domain knowledge (e.g., circuit topology, specification trade-offs, PVT variations, layout parasitics) for design robustness and leveraging BO to enhance RL's starting point for training efficiency. } We show that such a framework is superior in designing various analog circuits with higher accuracy, efficiency, and reliability.
We expect that our method would assist human designers in accelerating analog chip design with artificial agents that master massive circuitry optimization experiences via learning.

\footnotesize

\end{document}